\documentclass[aps,prc,twocolumn,showpacs,amsmath,amssymb,nofootinbib]{revtex4-1}
\usepackage{graphicx}
\usepackage{color}
\usepackage{times}
\usepackage{inputenc}
\usepackage{bm}
\usepackage{ulem}
\usepackage{multirow}
\usepackage{float}
\usepackage{url}
\usepackage{natbib}
\usepackage{mathrsfs}
\usepackage{physics}
\usepackage[table,xcdraw]{xcolor}
\usepackage{booktabs}
\usepackage[colorlinks=true,citecolor=blue,urlcolor=blue,linkcolor=red]{hyperref}
\begin{document}
\title{Crustal properties of a neutron star within an effective relativistic mean-field model}
\author{Vishal Parmar$^{1}$}
\email{physics.vishal01@gmail.com}
\author{H. C. Das$^{2,3}$}
\email{harish.d@iopb.res.in}
\author{Ankit Kumar$^{2,3}$}
\email{ankit.k@iopb.res.in}
\author{M. K. Sharma$^{1}$}
\author{S. K. Patra$^{2,3}$} 
\email{patra@iopb.res.in}
\affiliation{\it $^{1}$ School of Physics and Materials Science, Thapar Institute of Engineering and Technology, Patiala 147004, India}
\affiliation{\it $^{2}$Institute of Physics, Sachivalaya Marg, Bhubaneswar 751005, India}
\affiliation{\it $^{3}$Homi Bhabha National Institute, Training School Complex, Anushakti Nagar, Mumbai 400094, India}
\date{\today}
\begin{abstract}
We use the effective relativistic mean-field (E-RMF) model to study the crustal properties of the neutron star. The unified equations of state (EoS) are constructed using recently developed E-RMF parameter sets, such as FSUGarnet, IOPB-I, and G3. The outer crust composition is determined using the atomic mass evaluation 2020 data [\href{https://iopscience.iop.org/article/10.1088/1674-1137/abddb0}{Chinese Physics C {\bf 45}, 030002 (2021)}] along with the available Hartree-Fock-Bogoliubov mass models [\href{https://journals.aps.org/prc/abstract/10.1103/PhysRevC.88.024308}{Phys. Rev. C {\bf 88}, 024308 (2013)}] for neutron-rich nuclei. The structure of the inner crust is estimated by performing the compressible liquid drop model calculations using the same E-RMF functional as that for the uniform nuclear matter in the liquid core. Various neutron star properties such as mass-radius ($M-R$) relation, the moment of inertia ($I$), the fractional crustal moment of inertia ($I_{crust}/I$), mass ($M_{crust})$ and thickness ($l_{crust}$) of the crust are calculated with three unified EoSs. The crustal properties are found to be sensitive to the density-dependent symmetry energy and slope parameter advocating the importance of the unified treatment of neutron star EoS. The three unified EoSs, IOPB-I-U, FSUGarnet-U, and G3-U, reproduced the observational data obtained with different pulsars, NICER, and glitch activity and are found suitable for further description of the structure of the neutron star. 
\end{abstract}
\maketitle
\section{Introduction}
\label{intro}
In 1934, astronomers Baade and Zwicky, in their pioneering work, coined the term ``supernova" and hypothesized the existence of neutron stars \cite{Baade254, Baade259} which was discovered by Hewish {\it {\it et al.}} in 1968 \cite{Hewish_1968}.  The discovery of neutron stars revolutionized nuclear and astrophysics and unfolded a new era of science. Neutron stars are one of the densest and most compact astrophysical objects, and the remnant collapsed core of giant stars with mass $8-20\ M_\odot$ after supernovae explosions \cite{Couch_2017}. The internal structure of a typical cold nonaccreting neutron star can be divided into three distinct parts below its thin atmosphere:  two concentric inhomogeneous outer and inner crust followed by a dense homogeneous liquid core \cite{Haensel_2008, Haensel_2009, BKS_2015}. The neutron star remains in complete thermodynamic equilibrium against all possible interactions and in the lowest energy state at zero temperature. The outermost layer, the ``outer crust", consists of a body-centered cubic (BCC) lattice embedded in the sea of electrons, making it globally charge neutral. With increasing star depth,  more and more neutron-rich nuclei appear until the onset of the inner crust, where neutrons start dripping, owing to high-density \cite{Chamel_2016, Roca_2008, Pearson_2018}. The inner crust is marked by the assembly of the clusters formed by neutrons and protons along with the unbound neutrons making the neutron gas. The system is neutralized by the electron gas, which is distributed uniformly over the cluster and neutron gas \cite{Margaritis_2021, Oyamatsu_2007}. The clusters can have different shapes such as the sphere, slab, rods, etc., commonly known as ``nuclear pasta" \cite{Avancini_2008, Avancini_2009}  to reduce the energy of the cluster.   As the density increases, the size of the cluster in the inner crust increases, and at a density called transition density, the inhomogeneities disappear, and we enter the liquid core of the star, which consists of an admixture of neutrons and protons along with the leptons ensuring the charge neutrality and $\beta-$ equilibrium.

Determining the structure of the neutron star from the surface to interiors in a unified way is one of the principal problems in neutron star physics. Apart from a small region of the outer crust, the structure of the neutron star is mainly dependent on the equation of state (EoS). A substantial amount of research has been carried out in the last two decades to constrain the EoS based on many experimental and theoretical observations \cite{Fattoyev_2012, Abbott_2018, Tuhin_2018, Miller_2019, Raaijmakers_2019, Greif_2020}. The GW170817 event \cite{Abbott_2017, Abbott_2018} provides an upper limit on the tidal deformability while the massive pulsar such as PSR J0740+6620 \cite{Cromartie_2019} ,   PSR J0348+0432 \cite{Antoniadis_2013} and PSR J1614–2230 \cite{Demorest_2010} estimate that the neutron star mass should be greater than $2~M_\odot$. There are just a few EoSs which have been used to calculate the neutron star structure in the entire density range within a unified approach and satisfy the relevant constraints \cite{Dutra_2014}. The unified treatment of the neutron star is essential as various properties such as crust-core transition density, pressure, the crustal moment of inertia, etc., are very sensitive to the choice of EoS \cite{Dutra_2021}. These properties and the structure of the crust, which essentially depends on the subsaturation behavior of EoS, have a significant impact on the transport and thermodynamical properties of the neutron star.

In this work, we provide a unified treatment of the structure of the neutron star within the effective relativistic mean-field (E-RMF) approach using the cold catalyzed matter approximation (CCM). The CCM means that the star is in thermal and $\beta-$ equilibrium, valid for any non-accreting neutron star \cite{Miller_2019}. The E-RMF formalism is inspired by the effective field theory (EFT), where we do not have to worry about the renormalization problem as in the conventional RMF theory \cite{Arumugam_2004, Furnstahl_1997}. The effective Lagrangian is consistent with the underlying quantum chromodynamics (QCD) symmetries and contains infinite terms, and none can be dropped without the proper symmetry argument \cite{Estal_2001}. The E-RMF theory has been very successful in the last two decades and has been applied in various nuclear problems, which range from the properties of the nucleus to the structure of the neutron star \cite{Kumar_2017,Hcdas_2020, Hcdas_2021, Ankit_2020}.

We begin our calculations from the surface of the star with a density greater than $10^{-10}$ fm$^{-3}$ where all the atoms are completely ionized, and electrons form a degenerate Fermi gas. Below this density, the electrons are still bounded to the nuclei, and one can use generalized Thomas-Fermi (TF) theory to calculate the properties of this thin layer \cite{Feynman_1949, Haenel_2007}. The composition of the outer crust, which starts from the density of $10^{-10}$ fm$^{-3}$ until the onset of neutron drip, is calculated using the pioneering variational formalism proposed by Baym-Pethick-Sutherland (BPS) \cite{BPS_1971}. It considers that the ensemble of heavy nuclei may be represented by a single nucleus commonly known as the single-nucleus approximation \cite{Shen_2011}, thus giving a unique configuration for given thermodynamic conditions. The only input in the calculation of outer crust is the atomic mass evaluations. We have taken the values from the recently measured atomic mass evaluation (AME) 2020 mass table \cite{Huang_2021}, which is available up to isospin asymmetry of 0.3. Mass evaluations are not possible for more neutron-rich nuclei in the laboratory, so the need to use a mass model arises. For this, we use the nuclear mass model calculated from the Hartree-Fock-Bogoliubov (HFB ) \cite{Samyn_2002} method using the accurately calibrated Brussels-Montreal \cite{Eya_2017} energy-density functionals, such as, BSk14, BSk24, and BSk26 \cite{hfb14, hfb2426}. The HFB approach is a highly precise formalism used in various calculations concerning nuclear masses for the highly neutron-rich nuclei. 

The onset of neutron drip marks the beginning of the inner crust, which has an intricate structure making it a challenging problem. Different treatments of inner crust are available such as microscopic calculation pioneered by Negele and Vautherin \cite{NEGELE1973298} using the microscopic Hartree-Fock approach and subsequently modified by Baldo {\it et al.} \cite{Baldo_2006}, and Onsi {\it et al.} \cite{Onsi_2008} which uses the extended Thomas-Fermi (ETF) formulation. The microscopic calculations that specifically include the quantum nature are accurate but suffer from the fact that one needs to solve boundary value problems and do not allow the specific treatment of different terms such as surface or Coulomb energy. On the other hand, classical formalism such as the compressible liquid drop model (CLDM) \cite{Mackie_1977, Newton_2021} is computationally economical and avoids the choice of boundary conditions. The CLDM model is modified from the conventional semiempirical model by Baym-Bethe-Pethick \cite{Baym_1971} which incorporated the compressibility of nuclear matter, negative lattice Coulomb energy, and the suppression of surface tension by the neutron gas. The results of CLDM are known to be at par with those of ETF, and TF calculations \cite{Newton_2012}. It should be noted that the CLDM requires that the same functional be used for the calculation of bulk as well as the finite size contributions. The CLDM is recently applied in the work of Refs. \cite{Carreau_2019, Carreau_2020, carreau2020modeling} where the energy-density functional is taken in the form of meta-modeling, a technique developed to mimic the original relativistic or nonrelativistic functional using the isoscalar and the isovector energy of the EoS \cite{Margueron_2018} and for the Bayesian inference of neutron star crust properties \cite{Newton_2021}.  The meta-modeling reduces the computational difficulties when studying the statistical properties such as Bayesian inference to constrain the EoS.  Although this formalism reasonably imitates the EoS at low density but deviates at extremely low and high density, thereby estimating different neutron star results as the original EoS. We, therefore, use the technique developed by Carreau {\it {\it et al.}} \cite{code} and modify it to use the exact E-RMF formalism for the calculation of bulk and finite-size contribution of the cluster. This will preserve the underlying properties of a parameter that may otherwise be lost in the meta-modeling.

The aim of this paper is twofold: First, we develop three unified EoS, namely FSUGarnet-U, IOPB-I-U, and G3-U  with available core EOSs, such as FSUGarnet \cite{Chen_2014}, IOPB-I \cite{Kumar_2018}, and G3 \cite{Kumar_2017}. We construct the EoS from the outer crust to the liquid core using the experimental mass from the AME2020 data \cite{Huang_2021}, mass table of HFB-26 \cite{hfb2426}, available mass excess of neutron-rich nuclei \cite{wolf2013, welker2017, Beck_2021} and the E-RMF sets FSUGarnet \cite{Chen_2015}, IOPB-I \cite{Kumar_2018}, and G3 \cite{Kumar_2017}. We consider only spherical geometry for the estimation of inner crust structure. Second, we study the neutron star properties such as the $M-R$ relation, the moment of inertia. We study the influence of the crust on the moment of inertia in the form of fractional moment of inertia (FMI) which plays an important role to understand the pulsar glitch behavior \cite{Basu_2018, Eya_2017}. Pulsar glitches are the sudden jump in the spin frequency usually attributed to the depth of their interior superﬂuid from the surface. Therefore, these glitches are related to the crust thickness and act as the laboratory to test the validity of nuclear models.  

The paper is organized as follows: In Sec. \ref{formulaion}, we describe the formalism for the solid outer crust, inner crust, and liquid core of the neutron star. We, in brief, describe the E-RMF formalism and neutron star observables such as the moment of inertia. We discuss the results in Sec. \ref{results}. Finally, we summarize our results in Section \ref{conclusion}. 
\section{\label{formulaion} Formulation}
\subsection{Outer crust}
In the outer crust, the energy of Wigner–Seitz (WS) cell at a given baryon density ($\rho_b$) with the condition of charge neutrality is given by \cite{Haensel_2008}
\begin{equation}
    \label{wsenergy}
    E(A,Z,\rho_b)_{WS}= E(A,Z)_N + E_L+E_{zp}+E_e,
\end{equation}
where $E(A, Z)_N=M(A, Z)$ is the rest mass energy of nucleus with mass number $A$ and atomic number $Z$. $E_L$ and $E_{zp}$ corresponds to static-lattice and zero-point energy, which are written as \cite{carreau2020modeling}
\begin{equation}
    \begin{aligned}
    &E_L=-C_M\frac{(Ze)^2}{R_N}; \ \ R_N=\left(\frac{3}{4 \pi} \rho_N\right)^{1/3},\\
    &E_{zp}=\frac{3}{2}\hbar\omega_pu.
    \end{aligned}
\end{equation}
Here, $C_M=0.895929255682$ is the Mandelung constant, $u=0.51138$ is a constant for a BCC lattice \cite{Chamel_2016} and $\omega_p$ is the plasma frequency. $\rho_N$ is the neutron density. $E_e=\mathcal{E}_e V_{WS}$ is the energy of the surrounding relativistic electron gas. $V_{WS}$ is the volume of the WS cell.

In order to estimate the composition of the ground state of the outer crust, we use the BPS technique \cite{BPS_1971}. At a fixed pressure, we find a nucleus with the mass number $A$ and charge $Z$ that minimizes the Gibbs free energy \cite{BPS_1971},
\begin{equation}
    G(A,Z,P)=\frac{\mathcal{E}_{WS}+P}{\rho_b},
\end{equation}
where $\mathcal{E}_{WS}=E_{WS}/V_{WS}$ is the energy density of WS cell and $\rho_b=A/V_{WS}=\rho_N A$ is the baryon density. The advantage of taking pressure as an independent variable is that it increases monotonically while moving from the surface to the core. Thus discontinuity in density suggests the transition from one layer of the nucleus to another.  One also gets rid of the Maxwell construction \cite{Vishal_2021_jpg} to determine the transition pressure from one nucleus to another.

The pressure can be calculated from the first law of thermodynamics as \cite{Carreau_2020} 
\begin{equation}
    P=\rho_b^2\frac{\partial \mathcal{E}_{WS}/\rho_b}{\partial \rho_b}.
\end{equation}
Nucleons exert no pressure in the outer crust, and the total pressure can be written using Eq. (\ref{wsenergy}) as
\begin{equation}
\label{ocpress}
    P=\frac{1}{3}E_L\rho_N +\frac{1}{2}E_{zp}\rho_N+P_e.
\end{equation}
The Gibbs free energy to minimize thus becomes \cite{bcpm, Carreau_2020}
\begin{equation}
\label{eq:gibbsminimization}
    G(A,Z,P)=\frac{M(A,Z)}{A}+\frac{4}{3}\frac{E_L}{A}+\frac{1}{2}\frac{E_{zp}}{A}+\frac{Z}{A}\mu_e,
\end{equation}
where $\mu_e$ is the electron chemical potential. The only input in the calculation of outer crust is the nuclear mass table which can be taken from experiments \cite{Huang_2021} which are available for $I=(N-Z)/A \leq 0.3$. For the nuclear mass of more neutron-rich nuclei, we use microscopic HFB theoretical mass tables \cite{Samyn_2002}. The outer crust extends to the density where the chemical potential of neutrons exceeds its rest mass-energy. The neutron chemical potential utilizing the  condition of $\beta-$equilibrium $\mu_n=\mu_p+\mu_e$ can be simply written as
\begin{equation}
    \mu_n=G.
\end{equation}
\subsection{Inner crust}
As one moves deeper into the crust, the neutrons become less and less bound. At the transition density, the neutrons drip out of the nuclei and start filling the continuous energy spectrum. The dripped neutrons stay confined in the WS cell due to the large gravitational pressure. 
In the inner crust, the WS consists of a cluster surrounding ultrarelativistic electron gas and ambient neutron gas. The energy of this cluster can be written as \cite{Pearson_2018, BPS_1971}
\begin{equation}
\label{wsenergyic}
E_{WS}=M_i(A,Z)+E_e+V_{WS}(\mathcal{E}_g+\rho_gM_n),
\end{equation}
where $M_i(A,Z)$ is the mass of the cluster written as
\begin{equation}
    \label{clustermass}
    M_i(A,Z)=(A-Z)M_n + Z M_p + E_{cl}-V_{cl}(\mathcal{E}_g+\rho_g M_n),
\end{equation}
where $M_n$, and $M_p$ are the masses of neutron and proton respectively. ${\cal E}_g$, and $\rho_g$ are the energy density and density of the neutron gas respectively. We use the CLDM to determine the energy of the cluster which reads
\begin{equation}
    \label{ecluster}
    E_{cl}=E_{bulk}(\rho_0,I)A+E_{surf}+E_{curv}+E_{coul},
\end{equation}
where $E_{surf}$, $E_{curv}$, and $E_{coul}$ are surface, curvature and Coulomb energy respectively. In WS approximation, the Coulomb energy, which consists of lattice and finite-size correction, is written as \cite{carreau2020modeling}
\begin{equation}
    E_{col}=\frac{3}{20}\frac{e^2}{r_0}\eta_{col}A^{5/3}(1-I)^2,
\end{equation}
with
\begin{equation}
    \eta_{col}=1-\frac{3}{2}\lambda^{1/3}+\frac{1}{2}\lambda
\end{equation}
where $\lambda=\rho_e/\rho_{0,p}$ is the volume fraction with $\rho_{0,p}$ and $\rho_e$ are the proton and electron density inside the cluster respectively. Considering cluster to be spherical, the surface energy is defined as
\begin{equation}
  E_{surf} = 4\pi R_0^2A^{2/3}\sigma(I),\label{eq:esurf}
\end{equation}
where $R_0 = (4\pi \rho_0/3)^{-1/3}$ is related to the cluster density $\rho_0$, and
$\sigma(I)$ is the nuclear surface tension that depends on the isospin asymmetry of the cluster.  We use the parametrization of surface tension proposed by Ravenhall \textit{{\it {\it et al.}}}~\cite{Ravenhall1983} which is obtained by fitting Thomas-Fermi and Hartree-Fock numerical values as,
\begin{equation}
  \sigma(I) = \sigma_0\frac{2^{p+1} + b_s}{Y_p^{-p} + b_s + (1 -
  Y_p)^{-p}},\label{eq:sigma}
\end{equation}
where, $\sigma_0,p,b_s$ are the free parameters and $Y_p$ is the proton fraction inside the cluster. Similar to surface energy, the curvature energy plays an important part in describing the surface and is written as \cite{Newton_2012} 
\begin{equation}
  E_{curv} = 8\pi r_0A^{1/3}\sigma_c.\label{eq:ecurv}
\end{equation}
Here $\sigma_c$ is the curvature tension related to the surface tension
$\sigma$ as \cite{carreau2020modeling, Newton_2012},
\begin{equation}
  \sigma_c =
  \sigma\frac{\sigma_{0,c}}{\sigma_0}\alpha(\beta-Y_p),\label{eq:sigmac}
\end{equation}
with $\alpha=5.5$  and $\sigma_{0,c}, \beta$ are the parameters which  along with the $\sigma_0$ and $b_s$ needs to be fitted for a given EoS with the available experimental AME2020 mass table \cite{Huang_2021} at a fixed value of $p$. The equilibrium composition of inhomogeneous matter in the inner crust is obtained by minimizing the energy of WS cell per unit volume at a given baryon density ($\rho_b=\rho_n+\rho_p$), where $\rho_n$ and $\rho_p$ represent the neutron and proton density respectively. We use the variational method used in \cite{Carreau_2019,Newton_2012} where the Lagrange multipliers technique is used so that the auxiliary function to be minimized reads as \cite{Carreau_2019, Carreau_2020}
\begin{equation}
    \label{eq:auxillaryfunctio}
    \mathscr{F}(A,I,\rho_0,\rho_g,\rho_p)=\frac{E_{WS}}{V_{WS}}-\mu_b\rho_b,
\end{equation}
where $\mu_b$ is the baryonic chemical potential given by \cite{Carreau_2019}
\begin{equation}
    \label{baryonicchempot}
    \mu_b=\frac{2\rho_0\rho_p}{\rho_0(1-I)-2\rho_p}\frac{\partial (E_{cl}/A)}{\partial \rho_g}+\frac{d \mathcal{E}_g}{d\rho_g}.
\end{equation}
The chemical and mechanical equilibrium along with the Bayam virial theorem then transmute to the following set of coupled differential equations  \cite{Carreau_2020},
\begin{subequations}
\label{diffeq}
\begin{equation}
  \frac{\partial (E_{cl}/A)}{\partial A} = 0,\label{eq:ic1}
\end{equation}
\begin{equation}
  \frac{\rho_0^2}{A}\frac{\partial E_{cl}}{\partial \rho_0} = P_g,\label{eq:ic2}
\end{equation}
\begin{equation}
  \frac{E_{cl}}{A} + \frac{1-I}{A}\frac{\partial E_{cl}}{\partial I} +
  \frac{P_g}{\rho_0} = \mu_g,\label{eq:ic3}
\end{equation}
\begin{equation}
  \frac{2}{A}\left(\frac{\partial E_{cl}}{\partial I} -
  \frac{\rho_p}{1-I}\frac{\partial E_{cl}}{\partial \rho_p}\right)  = \mu_e(\rho_p),
  \label{eq:ic4}
\end{equation}
\end{subequations}
where $P_g$ is the gas pressure. The four differential equations (\ref{diffeq}) are solved simultaneously to estimate the equilibrium composition in the inner crust.  The energy density for the homogeneous nuclear matter entering Eq. (\ref{ecluster}) and neutron gas in this work is determined employing the effective relativistic mean-field theory, which will be discussed in the next section. 
\subsection{Liquid core}
As the density is increased, the transition from inner solid crust to outer liquid core takes place. In the outer core, the energy density of homogeneous matter is written as
\begin{equation}
    \label{coreenergy}
    \mathcal{E}_{core}=\mathcal{E}_{B}(\rho_b,\alpha)+\mathcal{E}_{e}(\rho_e)+\mathcal{E}_{\mu}(\rho_\mu),
\end{equation}
where $B$ stands for  baryon. The population of baryons and leptons are calculated by the constraints of $\beta-$equilibrium and charge neutrality as  \cite{NKGb_1997, Hcdas_2020,Hcdas_2021}
\begin{subequations}
\begin{equation}
\mu_n=\mu_p+\mu_e,  \ \ \mu_e=\mu_\mu.
\end{equation}
\begin{equation}
    \rho_p=\rho_e+\rho_\mu,
\end{equation}
\end{subequations}
where $\mu_{p,n,e,\mu}$ are the chemical potential of the proton, neutron electron, and muon in the homogeneous phase respectively. We define the crust-core transition from the crust side when the energy density of the WS cell in the inner crust exceeds the energy density of the liquid core. It can be written as
\begin{equation}
    \label{eq:cctransition}
    \mathcal{E}_{WS}({\rho_t})=\mathcal{E}_{npe\mu}(\rho_{\rho_t}).
\end{equation}
\vspace{0.2cm}
\subsection{Effective relativistic mean-field theory}
The E-RMF formalism is inspired by the effective field theory (EFT) motivated relativistic mean field formalism and is consistent with the underlying QCD symmetries. The conventional RMF models, such as nonlinear NL-type (NL1, NL2, NL-SH, NL3 etc.), consider only the higher-order self-couplings of sigma-mesons. These couplings help to reduce the incompressibility of nuclear matter to less than 300 MeV ($\sim 210-270$ MeV) \cite{Reinhard_1986, Reinhard_1989, Lalazissis_1997, LALAZISSIS_2009}. 

Although, these models predict the incompressibility well within the experimental data, other nuclear matter properties of these models, such as symmetry energy and its higher-order coefficients, do not fall in the accepted empirical or experimental range \cite{Dutra_2014}. In addition, EoS calculated from these models also do not satisfy the flow data due to their stiffness, which is one of the major drawbacks of these models. Consequently, these models estimate the mass and radius of the neutron star more than $2.5\ M_\odot$ and $\sim 14$ km, which doesn't satisfy the latest massive pulsars and NICER data, respectively \cite{Dutra_2016}. However, these models are known to predict finite nuclei properties in agreement with the  experimental data. Apart from the conventional NL-type RMF models, a few modified models have also been proposed that are still unable to satisfy experimental/observational data for nuclear and neutron star cases \cite{Dutra_2016}.

The E-RMF Lagrangian, on the other hand,  includes higher-order terms both for self and cross-couplings between different mesons ($\sigma, \omega, \rho, {\rm and} \ \delta$) \cite{Furnstahl_1997, Del_2001}. In our case, we take the interaction between different mesons up to 4th order except $\rho^4$ and $\delta^4$ (in G3 and IOPB-I cases). The G3 set contains the $\delta$ meson, which plays an important role in the high-density limit and is absent in the majority of RMF models. The predicted nuclear matter properties such as incompressibility (220--250 MeV), symmetry energy (30--35 MeV), and its slope parameter (40--80 MeV) by standard E-RMF forces ( e.g., G3, IOPB-I, FSUGarnet, etc. ) are in agreement with different empirical/experimental data. The flow data constraint is also well satisfied by modern E-RMF sets \cite{Kumar_2017, Kumar_2018}. The most important point is that almost all the modern E-RMF parameter sets satisfy the $2 \ M_\odot$ constraint of neutron star. The E-RMF has the advantage that besides being excellent for calculating neutron star properties, it does not violate the predictive power of finite nuclei \cite{Kumar_2017, Kumar_2018}. Therefore, the E-RMF formalism is  as good as the conventional RMF framework and, in some cases, even performs better. This formalism has been applied in a wide range of nuclear physics problems in the past few years \cite{MULLER_1996, Wang_2000, DelPairing_2001, Kumar_2020, Das_2020, Das_2021, QuddusDM_2020}. The E-RMF effective Lagrangian which include the interaction between different mesons, such as, $\sigma$, $\omega$, $\rho $, $\delta$ and photon is written as \cite{Patra_2002, Kumar_2017, Kumar_2018, Vishal_2021_jpg, Vishal_2021, DasBig_2020}, 

\begin{widetext}
\begin{eqnarray}
\label{rmftlagrangian}
\mathcal{E}(r)&=&\psi^{\dagger}(r)\qty{i\alpha\cdot\grad+\beta[M-\Phi(r)-\tau_3D(r)]+W(r)+\frac{1}{2}\tau_3R(r)+\frac{1+\tau_3}{2} A(r)-\frac{i\beta \alpha }{2M}\qty(f_\omega \grad W(r)+\frac{1}{2}f_\rho \tau_3 \grad R(r))}\psi(r) \nonumber \\
&+& \qty(\frac{1}{2}+\frac{k_3\Phi(r)}{3!M}+\frac{k_4}{4!}\frac{\Phi^2(r)}{M^2})\frac{m^2_s}{g^2_s}\Phi(r)^2+\frac{1}{2g^2_s}\qty\Big(1+\alpha_1\frac{\Phi(r)}{M})(\grad \Phi(r))^2-\frac{1}{2g^2_\omega}\qty\Big(1+\alpha_2\frac{\Phi(r)}{M})(\grad W(r))^2 \nonumber\\
&-&\frac{1}{2}\qty\Big(1+\eta_1\frac{\Phi(r)}{M}+\frac{\eta_2}{2}\frac{\Phi^2(r)}{M^2})\frac{m^2_\omega}{g^2_\omega}W^2(r)-\frac{1}{2e^2}(\grad A^2(r))^2 -\frac{1}{2g^2_\rho}(\grad R(r))^2
-\frac{1}{2}\qty\Big(1+\eta_\rho\frac{\Phi(r)}{M})\frac{m^2_\rho}{g^2_\rho}R^2(r)\nonumber \\
&-&\frac{\zeta_0}{4!}\frac{1}{g^2_\omega}W(r)^4-\Lambda_\omega(R^2(r)W^2(r))
+\frac{1}{2g^2_\delta}(\grad D(r))^2
+\frac{1}{2}\frac{m^2_\delta}{g^2_\delta}(D(r))^2.
\end{eqnarray} 
\end{widetext}
Here $\Phi(r)$, W(r), R(r), D(r) and A(r) are the fields corresponding to $\sigma$, $\omega$, $\rho$ and 
$\delta $ mesons and photon respectively. The $g_s$, $g_{\omega}$, $g_{\rho}$, $g_{\delta}$ and $\frac{e^2}{4\pi }$ 
are the corresponding coupling constants and $m_s$, $m_{\omega}$, $m_{\rho}$ and $m_{\delta}$ are the 
corresponding masses. 
The  zeroth component $T_{00}= H$ and the third component $T_{ii}$ of energy-momentum tensor 
\begin{equation}
\label{set}
T_{\mu\nu}=\partial^\nu\phi(x))\frac{\partial\mathcal{E}}{\partial\partial_\mu \phi(x)}-\eta^{\nu\mu}\mathcal{E},
\end{equation}
yields the energy and pressure density, respectively as \cite{Hcdas_2020, Das_2021, Kumar_2020}
\begin{eqnarray}
{\cal{E}}&=& \frac{\gamma}{(2\pi)^{3}}\sum_{i=p,n}\int_0^{k_i} d^{3}k E_{i}^\star (k_i)+\rho_bW +\frac{1}{2}\rho_{3}R 
\nonumber\\
&+&\frac{ m_{s}^2\Phi^{2}}{g_{s}^2}\Bigg(\frac{1}{2}+\frac{\kappa_{3}}{3!}
\frac{\Phi }{M_{nucl.}} + \frac{\kappa_4}{4!}\frac{\Phi^2}{M_{nucl.}^2}\Bigg) -\frac{1}{4!}\frac{\zeta_{0}W^{4}}{g_{\omega}^2}
\nonumber\\
&-&\frac{1}{2}m_{\omega}^2\frac{W^{2}}{g_{\omega}^2}\Bigg(1+\eta_{1}\frac{\Phi}{M_{nucl.}}+\frac{\eta_{2}}{2}\frac{\Phi ^2}{M_{nucl.}^2}\Bigg)
\nonumber\\
&-&\Lambda_{\omega}  (R^{2}\times W^{2})-\frac{1}{2}\Bigg(1+\frac{\eta_{\rho}\Phi}{M_{nucl.}}\Bigg)\frac{m_{\rho}^2}{g_{\rho}^2}R^{2}
\nonumber\\
&+&\frac{1}{2}\frac{m_{\delta}^2}{g_{\delta}^{2}}D^{2},
\label{eq:energy}
\end{eqnarray}
\noindent
\begin{eqnarray}
P&=&  \frac{\gamma}{3 (2\pi)^{3}}\sum_{i=p,n}\int_0^{k_i} d^{3}k \frac{k^2}{E_{i}^\star (k_i)}+\frac{1}{4!}\frac{\zeta_{0}W^{4}}{g_{\omega}^2}
\nonumber\\
&&
-\frac{ m_{s}^2\Phi^{2}}{g_{s}^2}\Bigg(\frac{1}{2}+\frac{\kappa_{3}}{3!}
\frac{\Phi }{M_{nucl.}}+ \frac{\kappa_4}{4!}\frac{\Phi^2}{M_{nucl.}^2}\Bigg)
\nonumber\\
&&
+\frac{1}{2}m_{\omega}^2\frac{W^{2}}{g_{\omega}^2}\Bigg(1+\eta_{1}\frac{\Phi}{M_{nucl.}}+\frac{\eta_{2}}{2}\frac{\Phi ^2}{M_{nucl.}^2}\Bigg)
\nonumber\\
&&
+\Lambda_{\omega} (R^{2}\times W^{2})+\frac{1}{2}\Bigg(1+\frac{\eta_{\rho}\Phi}{M_{nucl.}}\Bigg)\frac{m_{\rho}^2}{g_{\rho}^2}R^{2}
\nonumber\\
&&
-\frac{1}{2}\frac{m_{\delta}^2}{g_{\delta}^{2}}D^{2}.
\label{eq:press}
\end{eqnarray}
We tabulated the different masses of the mesons and coupling constants in Table \ref{table1} for three considered parameter sets FSUGarnet, IOPB-I and G3.
\begin{table} 
\centering
\caption{The masses and coupling constants for FSUGarnet \cite{Chen_2015}, IOPB-I \cite{Kumar_2018}, and G3 \cite{Kumar_2017} are listed. The mass of nucleon $M$ is 939 MeV and other coupling constants are dimensionless.}
\begin{tabular}{ccccccccc}
\hline
\hline
\multicolumn{1}{c}{Parameter}
&\multicolumn{1}{c}{FSUGarnet}
&\multicolumn{1}{c}{G3}
&\multicolumn{1}{c}{IOPB-I}\\
\hline
$m_{s}/M$ &  0.529&  0.559&0.533\\
$m_{\omega}/M$& 0.833 &  0.832&0.833 \\
$m_{\rho}/M$& 0.812 &  0.820&0.812\\
$m_{\delta}/M$&  0.0&   1.043&0.0\\
$g_{s}/4 \pi$ &  0.837 &  0.782 &0.827 \\
$g_{\omega}/4\pi$ & 1.091 &  0.923&1.062 \\
$g_{\rho}/4 \pi$ & 1.105&  0.962 &0.885 \\
$g_{\delta}/4 \pi$&  0.0&  0.160& 0.0 \\
$k_{3} $ & 1.368&    2.606 &1.496\\
$k_{4}$ &  -1.397& 1.694 &-2.932  \\
$\zeta_{0}$& 4.410&  1.010  &3.103 \\
$\eta_{1}$ & 0.0&  0.424 &0.0\\
$\eta_{2}$ & 0.0&  0.114 &0.0\\
$\eta_{\rho}$& 0.0&  0.645& 0.0  \\
$\Lambda_{\omega}$ &0.043 &  0.038&0.024 \\
$\alpha_{1}$  & 0.0&   2.000&0.0 \\
$\alpha_{2}$  & 0.0&  -1.468&0.0  \\
$f_\omega/4$  & 0.0&  0.220&0.0\\
$f_\rho/4$ & 0.0&    1.239&0.0\\
$\beta_\sigma$ & 0.0& -0.087& 0.0\\
$\beta_\omega$  & 0.0& -0.484& 0.0 \\
\hline
\hline
\end{tabular}
\label{table1}
\end{table}
\subsection{Neutron star observables}
\label{form:NS_observable}
The metric corresponds to static, spherically symmetric stars is in the form of
\begin{eqnarray}
ds^2= -e^{2\nu(r)}dt^2+e^{2\lambda(r)}dr^2+r^2d\theta^2+r^2sin^2\theta d\phi^2,
\label{eq:metric}
\end{eqnarray}
where $r$, $\theta$ and $\phi$ are the coordinates. $\nu(r)$, $\lambda(r)$ are the metric potential given as  \citep{Krastev_2008}
\begin{eqnarray}
e^{2\lambda(r)} = [1-\gamma(r)]^{-1},
\end{eqnarray}
\begin{eqnarray}
e^{2\nu(r)}&=&e^{-2\lambda(r)} = [1-\gamma(r)], \qquad r>R_{star}
\end{eqnarray}
with
\begin{equation}
\gamma(r)=\left\{
\begin{array}{l l}
\frac{2m(r)}{r}, & \quad \mbox{if $r<R_{star}$}\\\\
\frac{2M}{R}, & \quad \mbox{if $r>R_{star}$}
\end{array}
\right.
\end{equation}
For static star, its macroscopic properties such as $M$ and $R$ of the NS, one can find by solving the Tolmann-Oppenheimer-Volkoff equations as follow \cite{TOV1, TOV2}
\begin{eqnarray}
\frac{dP(r)}{dr}= - \frac{[P(r)+{\cal{E}}(r)][m(r)+4\pi r^3 P(r)]}{r[r-2m(r)]},
\label{eq:pr}
\end{eqnarray}
and 
\begin{eqnarray}
\frac{dm(r)}{dr}=4\pi r^2 {\cal{E}}(r).
\label{eq:mr}
\end{eqnarray}
The $M$ and $R$ of the star can be calculated with boundary conditions $r=0, P = P_c$ and $r=R, P = P_0$ at certain central density.
    
The metric of slowly, uniformly rotating NS is given by \citep{Stergioulas_2003} 
\begin{eqnarray}
ds^2= -e^{2\nu}dt^2+e^{2\psi}(d\phi - \omega dt^2)+e^{2\alpha}(r^2d\theta^2+ d\phi^2),
\end{eqnarray}

The moment of inertia (MI) of the NS is calculated in the Refs.  \cite{Stergioulas_2003,Jha_2008,Sharma_2009,Friedmanstergioulas_2013,Paschalidis_2017,Quddus_2019,Koliogiannis_2020}. The expression of $I$ of uniformly rotating NS with angular frequency $\omega$ is given as \cite{Hartle_1967,Lattimer_2000,Worley_2008}
\begin{equation}
I \approx \frac{8\pi}{3}\int_{0}^{R}\ dr \ ({\cal E}+P)\  e^{-\phi(r)}\Big[1-\frac{2m(r)}{r}\Big]^{-1}\frac{\Bar{\omega}}{\Omega}\ r^4,
\label{eq:moi}
\end{equation}
where $\Bar{\omega}$ is the dragging angular velocity for a uniformly rotating star. The $\Bar{\omega}$ satisfying the boundary conditions are 
\begin{equation}
\Bar{\omega}(r=R)=1-\frac{2I}{R^3},\qquad \frac{d\Bar{\omega}}{dr}\Big|_{r=0}=0 .
\label{eq:omegabar}
\end{equation}
We calculate the crustal MI by using the Eq. (\ref{eq:moi}) from transition radius ($R_c$) to the surface of the star ($R$) is given by \cite{Fattoyev_2010, Basu_2018} 
\begin{equation}
I_{crust} \approx \frac{8\pi}{3}\int_{R_c}^{R}\ dr\ ({\cal E}+P)\  e^{-\phi(r)}\Big[1-\frac{2m(r)}{r}\Big]^{-1}\frac{\Bar{\omega}}{\Omega}\ r^4.
\label{eq:moic}
\end{equation}
\section{\label{results} Results and Discussions}
\subsection{Outer crust}
In the outer crust of the cold nonaccreting neutron star, the neutron-rich nuclei are embedded in a BCC lattice arrangement, ensuring that the cell's Coulomb energy is minimized. These nuclei are stable against the $\beta-$decay by surrounding uniform relativistic electron gas. To calculate the composition of the outer crust of a neutron star, we minimize the Gibbs free energy in Eq. (\ref{eq:gibbsminimization}) at fixed pressure where the atomic mass table serves as an input. We use the most recent AME2020 data \cite{Huang_2021} along with the recently measured mass excess of $^{77-79}$ Cu taken from \cite{welker2017}, $^{82}$Zn from \cite{wolf2013} and $^{151-157}$Yb \cite{Beck_2021}  for the known masses and extrapolate them using the microscopic HFB calculation namely HFB-24, HFB-26 \cite{hfb2426}, and HFB-14 \cite{hfb14},  which are based on BSk functional  characterized by unconventional Skyrme forces along with the most recent FRDM(2012) \cite{MOLLER20161} mass table.
\begin{figure}
    \centering
    \includegraphics[width=0.45\textwidth]{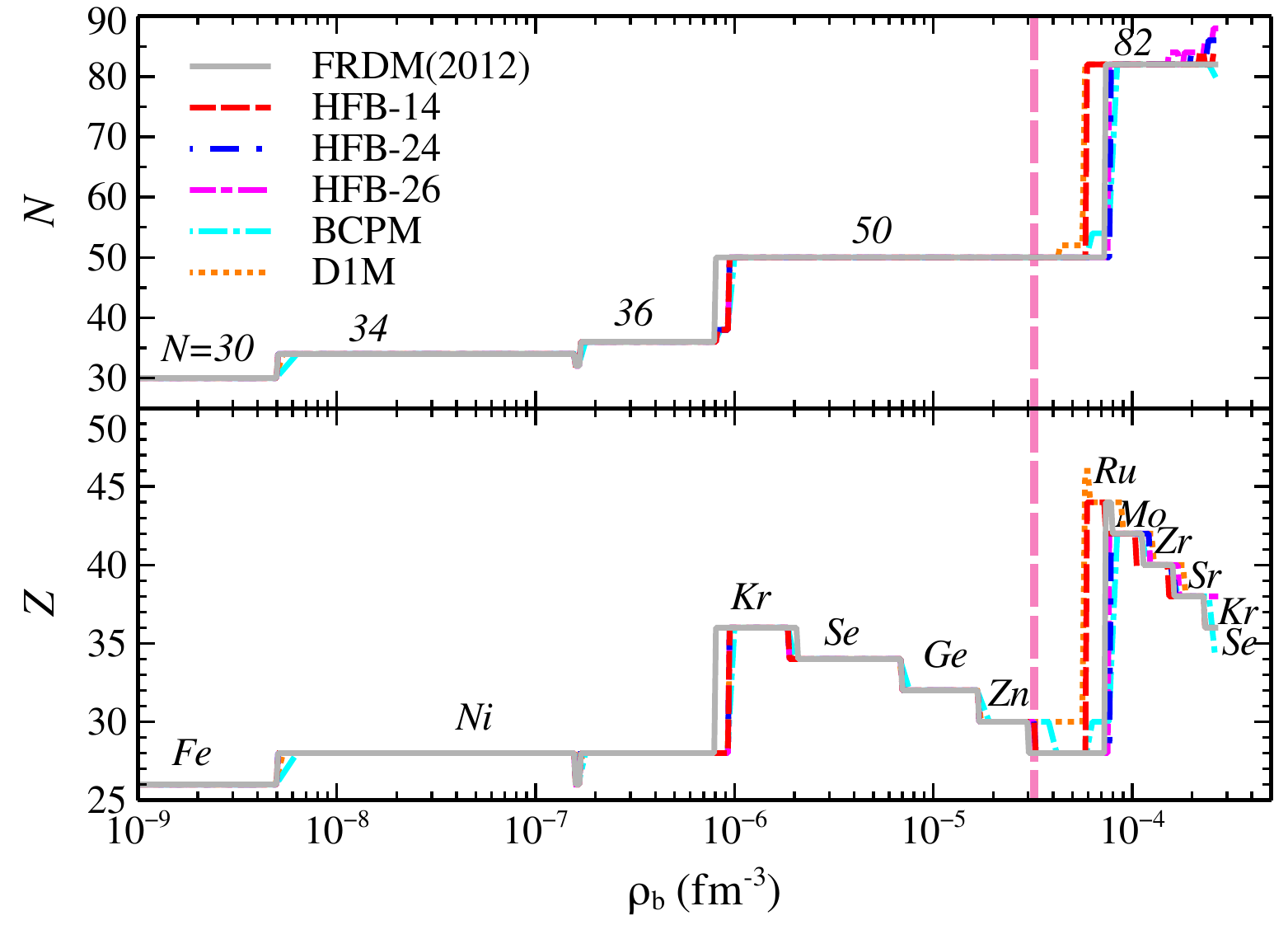}
    \caption{The proton ($Z$) and neutron number ($N$) in the outer crust as a function of density. The experimental data are taken from AME2020 when available \cite{Huang_2021}. The unknown mass are taken from microscopic calculations HFB-14 \cite{hfb14}, HFB-24 , HFB-26 \cite{hfb2426} along with the FRDM(2012) mass table \cite{MOLLER20161}. A comparison with BCPM  \cite{BKS_2015} and D1M \cite{sym13091613} is also shown. In addition the experimental mass of $^{82}$Zn \cite{wolf2013}, $^{77-79}$Cu \cite{welker2017} and $^{151-157}$Yb \cite{Beck_2021} are also considered. Vertical dashed line represent the boundary where prediction from experimental masses ends. }
    \label{fig:zndistribution}
\end{figure}

The composition of outer crust as a function of average baryon density is shown in Fig. \ref{fig:zndistribution} for the various mass models. In addition to the HFB computed mass excess, we also show the result from most recent FRDM(2012) \cite{MOLLER20161}, BCPM \cite{BKS_2015} and D1M \cite{sym13091613} Gogny interaction for a comparative analysis. The outermost layer is occupied by the $^{56}$Fe nucleus accompanied by the layer of $^{28}$Ni nucleus in the intermediate densities. The persistent existence of nuclear magic shell nuclei is also visible in $Z=28$ and $N=50$, $82$ plateau due to their enhanced binding energies. The layer of $N=50$ starts at density $\approx 10^{-6}$ fm$^{-3}$ and is characterized by the staircase structure signifying the decrease in atomic number due to the electron capture process. It leads to the appearance of more and more neutron-rich nuclei once we move deeper into the crust. The composition of the outer crust is determined solely from the experimental mass table up to the density $3.2\times 10^{-5}$ fm$^{-3}$ for the HFB-26, which is marked by the dashed vertical line in Fig. \ref{fig:zndistribution}. The composition is the model-independent until this density which is clear from the fact that all the curves overlap each other. It may be noted that the value of this density is slightly lower than the value determined from the AME2016 data.
\begin{figure}
    \centering
    \includegraphics[width=0.45\textwidth]{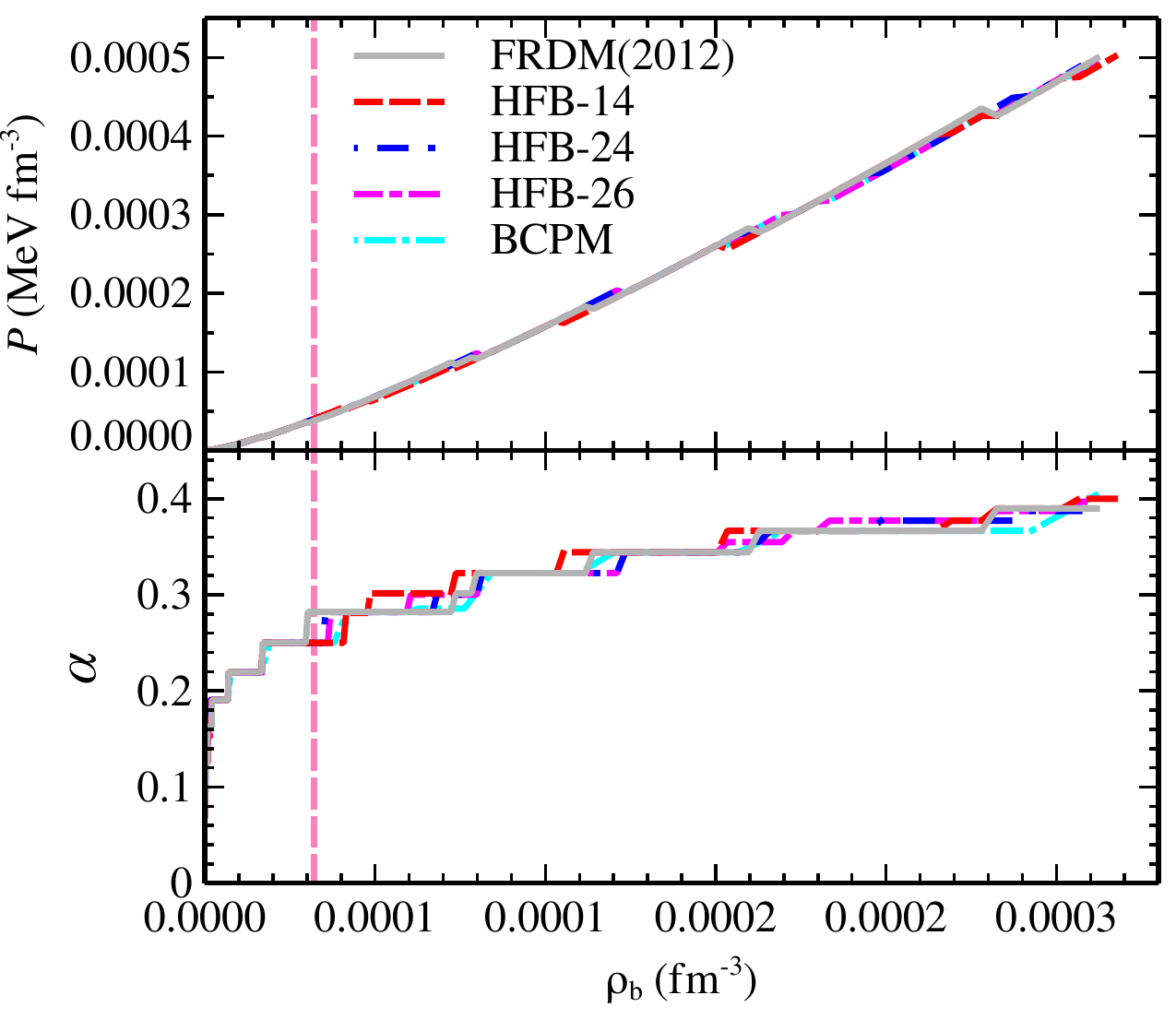}
    \caption{In the upper panel the EoS of outer crust is shown for different mass model. The lower panel shows the global asymmetry as a function of density. Vertical dashed line represent the boundary where prediction from experimental masses ends. }
    \label{fig:eosoc}
\end{figure}

As we move deeper into the outer crust, the need to apply a mass model to calculate the mass excess of extremely neutron-rich nuclei arises as these values are difficult to obtain in a laboratory setup. However, various advanced radioactive beam facilities are working toward measuring the properties of these neutron-rich nuclei in order to have a better understanding of the unconventional regime \cite{Dilling_2006, Beck_2021}. The highly precise HFB calculations and those obtained from the FRDM(2012), BCPM, and D1M predict the appearance of the $N=82$ layer at high density (near the transition to the inner crust), which is also marked by the staircaselike structure.  However, the model dependency is clearly visible in this case. The HFB calculations using HFB-14, HFB-24, and HFB-26 are close to the calculation of highly successful FRDM. For comparison of different models, we show the last two layers of the outer crust in Table \ref{tab:twonuc} where the last element corresponds to the layer just before the transition into the inner crust. In the entire outer crust, one can see a strong effect of closed proton and neutron shells on the composition, except for the outermost layer of $^{56}$Fe nucleus. The existence of nuclei with $Z=28$ and $N=50$ is the consequence of experimental fact whereas, $N=82$ can be treated as the artifact of extrapolation via the microscopic mass table used. In addition to these, there appears a thin layer of $^{121}$Y at the density 0.0001596 fm$^{-3}$ using the HFB-24 mass model. The existence of an odd mass or charge number in the outer crust is not considered in the  calculations of BPS \cite{BPS_1971} and signifies a possible ferromagnetic phase transition in a neutron star. Although one needs a more precise evaluation of the mass of odd-nuclei as it can alter the composition \cite{Pearson_2018} of the outer crust.
\begin{table}
\label{tab:twonuc}
\centering
\caption{The last two layers of nucleus in the outer crust predicted from the different model.}
\scalebox{0.92}{
\begin{tabular}{llllllll}
\hline
\hline
Model & Element & $Z$  & $N$  &
\begin{tabular}[c]{@{}l@{}}$ \hfill \rho_{max}$\\ (fm$^{-3}$)\end{tabular}  &
\begin{tabular}[c]{@{}l@{}} \hspace{0.5cm} $P$\\(MeV fm$^{-3}$) \end{tabular} &
\begin{tabular}[c]{@{}l@{}} \hspace{0.5cm} ${\cal E}$\\(MeV fm$^{-3}$) \end{tabular} & \hspace{0.2cm} $\alpha$
\\\hline
\multirow{2}{*}{HFB-14}    &   $^{122}$Sr      & 38 & 84 & 2.2799E-04 & 4.2566E-04 & 0.2137 & 0.377 \\
                           &   $^{120}$Kr      & 36 & 84 & 2.6712E-04 & 5.0108E-04 & 0.2505 & 0.400 \\
\hline                           
\multirow{2}{*}{HFB-24}    &    $^{122}$Sr     & 38 & 84 & 2.3720E-04 & 4.4874E-04 & 0.2224 & 0.377 \\
                           &    $^{124}$Sr     & 38 & 86 & 2.5675E-04 & 4.8804E-04 & 0.2407 & 0.387 \\
\hline                           
\multirow{2}{*}{HFB-26}    &    $^{122}$Sr     & 38 & 84 & 2.2799E-04 & 4.2566E-04 & 0.2137 & 0.377 \\
                           &    $^{126}$Sr     & 38 & 88 & 2.6188E-04 & 4.9052E-04 & 0.2456 & 0.397 \\
\hline                           
\multirow{2}{*}{FRDM}      &$^{120}$Sr         & 38 & 82 & 2.2799E-04 & 4.3515E-04 & 0.2137 & 0.367 \\
                           &$^{118}$Kr         & 36 & 82 & 2.6188E-04 & 4.9909E-04 & 0.2456 & 0.390 \\
\hline                           
\multirow{2}{*}{BCPM}      &$^{120}$Sr         & 38 & 82 & 2.4265E-04 & 4.7276E-04 & 0.2275 & 0.367 \\
                           &$^{114}$Se         & 34 & 80 & 2.6155E-04 & 4.8422E-04 & 0.2453 & 0.404 \\
\hline                           
\multirow{2}{*}{D1M}      & $^{122}$Zr        & 40 & 82 & 1.7990E-04 & 3.3165E-04 &       0.1685 & 0.344 \\
                           &$^{120}$Sr         & 38 & 82 & 2.4420E-04 & 4.7680E-04 &  0.2289      & 0.367 \\
\hline  \hline
\end{tabular}%
}
\end{table}

In Fig. \ref{fig:eosoc} we have shown the equation of state and the variation of global isospin asymmetry in the outer crust and tabulated data for HFB-26 in Table \ref{tab:oceosdata}.  The outer crust is marked by the discontinuous transition in the density at some pressure values, indicating a change of equilibrium nucleus. The pressure and chemical potential remain constant during the transition from one nucleus to another resulting in the finite shift in baryon density of the system. However, it is shown in Ref.  \cite{Jog_1982} that the transition between one layer to another layer takes place through a thin layer of the mixed state of two species with a pressure interval of $\approx$ 10$^{-4} P$. It should be noted here that the pressure of the outer crust is mainly determined from the relativistic electron gas as suggested in Eq. (\ref{ocpress}). The HFB calculations estimate similar EoS for the outer crust except at the points where the transition in the nucleus layers takes place. One can see that the majority of the outer crust is determined from the nuclear mass models, which are used to calculate the mass excess of neutron-rich nuclei. The inner layers of heavy nuclei account for the maximum mass of the outer crust. We also notice that the asymmetry increases monotonically with density, although relatively at a slower pace at high density in the outer crust, reaching $\approx 0.4$ at the transition from outer to the inner crust. The relative difference among different HFB mass models is also visible, attributed to their different symmetry energy. The symmetry energy plays a prominent role in determining the outer and inner crust structure and will be discussed in the next section.  
\begin{table}
\centering
\caption{The composition and EoS of outer crust. The experimental atomic mass evaluations are taken from AME2020 \cite{Huang_2021} when available. The unknown mass are taken from microscopic calculations HFB-26 \cite{hfb2426} .  In addition the experimental mass of $^{82}$Zn \cite{wolf2013}, $^{77-79}$Cu \cite{welker2017}  and $^{151-157}$Yb \cite{Beck_2021} are also considered. Horizontal  solid line represents the boundary where prediction from experimental masses ends. The upper part is obtained from the experimental data and the lower part from the HFB-26 results.}
\label{tab:oceosdata}
\renewcommand{\tabcolsep}{0.3cm}
\renewcommand{\arraystretch}{1.0}
\begin{tabular}{lllll}
\hline
\hline
\begin{tabular}[c]{@{}l@{}} \hspace{0.2cm} $ \rho_b$ \\(fm$^{-3}$) \end{tabular} &
\begin{tabular}[c]{@{}l@{}} \hspace{0.5cm} $P$\\(MeV fm$^{-3}$) \end{tabular}  &
\begin{tabular}[c]{@{}l@{}} \hspace{0.5cm} ${\cal E}$\\(MeV fm$^{-3}$) \end{tabular} & $Z$ & $N$ \\
\hline
1.0000E-09 & 2.9973E-11 & 9.3046E-07 & 26 & 30 \\
4.9730E-09 & 3.4018E-10 & 4.6275E-06 & 26 & 30 \\
5.0724E-09 & 3.3533E-10 & 4.7201E-06 & 28 & 34 \\
1.5597E-07 & 4.0911E-08 & 1.4522E-04 & 28 & 34 \\
1.5909E-07 & 4.1697E-08 & 1.4812E-04 & 26 & 32 \\
1.6552E-07 & 4.3999E-08 & 1.5411E-04 & 26 & 32 \\
1.6883E-07 & 4.3634E-08 & 1.5719E-04 & 28 & 36 \\
8.0697E-07 & 3.5983E-07 & 7.5177E-04 & 28 & 36 \\
8.2311E-07 & 3.5457E-07 & 7.6682E-04 & 28 & 38 \\
9.2696E-07 & 4.1587E-07 & 8.6361E-04 & 28 & 38 \\
9.4550E-07 & 4.1607E-07 & 8.8089E-04 & 36 & 50 \\
1.8538E-06 & 1.0258E-06 & 1.7278E-03 & 36 & 50 \\
1.8909E-06 & 1.0090E-06 & 1.7623E-03 & 34 & 50 \\
6.8498E-06 & 5.6411E-06 & 6.3900E-03 & 34 & 50 \\
6.9868E-06 & 5.5275E-06 & 6.5179E-03 & 32 & 50 \\
1.6699E-05 & 1.7692E-05 & 1.5592E-02 & 32 & 50 \\
1.7033E-05 & 1.7260E-05 & 1.5904E-02 & 30 & 50 \\
3.2099E-05 & 4.0208E-05 & 2.9994E-02 & 30 & 50 \\
\\
3.2741E-05 & 3.9028E-05 & 3.0595E-02 & 28 & 50 \\
7.5214E-05 & 1.1838E-04 & 7.0370E-02 & 28 & 50 \\
7.6718E-05 & 1.1094E-04 & 7.1779E-02 & 42 & 82 \\
1.2098E-04 & 2.0367E-04 & 1.1328E-01 & 42 & 82 \\
1.2340E-04 & 2.0062E-04 & 1.1554E-01 & 40 & 82 \\
1.5042E-04 & 2.6126E-04 & 1.4090E-01 & 40 & 82 \\
1.5343E-04 & 2.6250E-04 & 1.4372E-01 & 40 & 84 \\
1.6940E-04 & 2.9956E-04 & 1.5871E-01 & 40 & 84 \\
1.7278E-04 & 3.0065E-04 & 1.6189E-01 & 38 & 82 \\
1.7624E-04 & 3.0869E-04 & 1.6513E-01 & 38 & 82 \\
1.7977E-04 & 3.1695E-04 & 1.6844E-01 & 38 & 82 \\
1.8336E-04 & 3.1834E-04 & 1.7182E-01 & 38 & 84 \\
2.2799E-04 & 4.2566E-04 & 2.1372E-01 & 38 & 84 \\
2.3255E-04 & 4.2767E-04 & 2.1801E-01 & 38 & 86 \\
2.5171E-04 & 4.7532E-04 & 2.3601E-01 & 38 & 86 \\
2.5675E-04 & 4.7774E-04 & 2.4074E-01 & 38 & 88 \\
2.6188E-04 & 4.9052E-04 & 2.4557E-01 & 38 & 88 \\
\hline \hline
\end{tabular}%
\end{table}
\subsection{Inner crust}
With the increase in density or the distance from the star's surface, neutron chemical potential increases monotonically. When the chemical potential exceeds the rest mass of the neutron, the neutron starts dripping out of nuclei making the onset of the inner crust. Since no such system can be produced in terrestrial laboratories as neutrons evaporate, the inner crust inevitably becomes model dependent. We use the E-RMF model to calculate the properties of the inner crust using three recently developed parameter sets, namely FSUGarnet \cite{Chen_2014}, IOPB-I \cite{Kumar_2018}, and G3 \cite{Kumar_2017}. The bulk properties of these three parameter sets are provided in Table \ref{bulkproperties} along with the theoretical or experimental constraints. 
\begin{table}
\centering
\caption{Bulk matter properties such as saturation density ($\rho_{\rm sat}$), binding energy ($E_0$), effective mass ($m^*$), symmetry energy ($J$), slope parameter ($L$), second ($K_{sym}$) and third ($Q_{sym}$) order derivative of symmetry energy , incompressibility ($K$) of nuclear matter for the NL3, FSUGarnet, IOPB-I and G3 parameter and their corresponding empirical values}

\begin{tabular*}{\linewidth}{c @{\extracolsep{\fill}}ccccccc}
\hline
\hline
&NL3 & IOPB-I & G3 & FSUGarnet&Empirical Value &  \\ \hline
$\rho_{\rm sat}$ (fm$^{-3})$&0.148 & 0.149       &0.148 &0.153    &    0.148/0.185  \cite{bethe}     &  \\
$E_0$ (MeV) & -16.29 & -16.10  &-16.02   &-16.23 &   -15.0/-17.0   \cite{bethe} && \\
$M*/M$  & 0.595&0.593  &  0.699 &0.578 &  0.55/0.6 \cite{marketin2007}&\\
$J$ (MeV)& 37.43 &33.30  & 31.84  &30.95 &30.0/33.70 \cite{DANIELEWICZ20141}&  \\
$L$ (MeV) & 118.65  &63.58  & 49.31  &51.04 &35.0/70.0  \cite{DANIELEWICZ20141}&  \\
$K_{sym}$ (MeV)&101.34 & -37.09&-106.07&59.36& -174.0/31.0 \cite{zimmerman2020measuring}            &  \\
$Q _{sym}$ (MeV) &177.90& 862.70 &915.47 &130.93 & -494/-10 \cite{cai2017constraints}            &  \\
$K$ (MeV)  &271.38 & 222.65 &243.96 &229.5 &  220/260 \cite{GARG201855} &  \\
\hline     
\hline     
\end{tabular*}
\label{bulkproperties}
\end{table}

\begin{figure}
    \centering
    \includegraphics[width=0.45\textwidth]{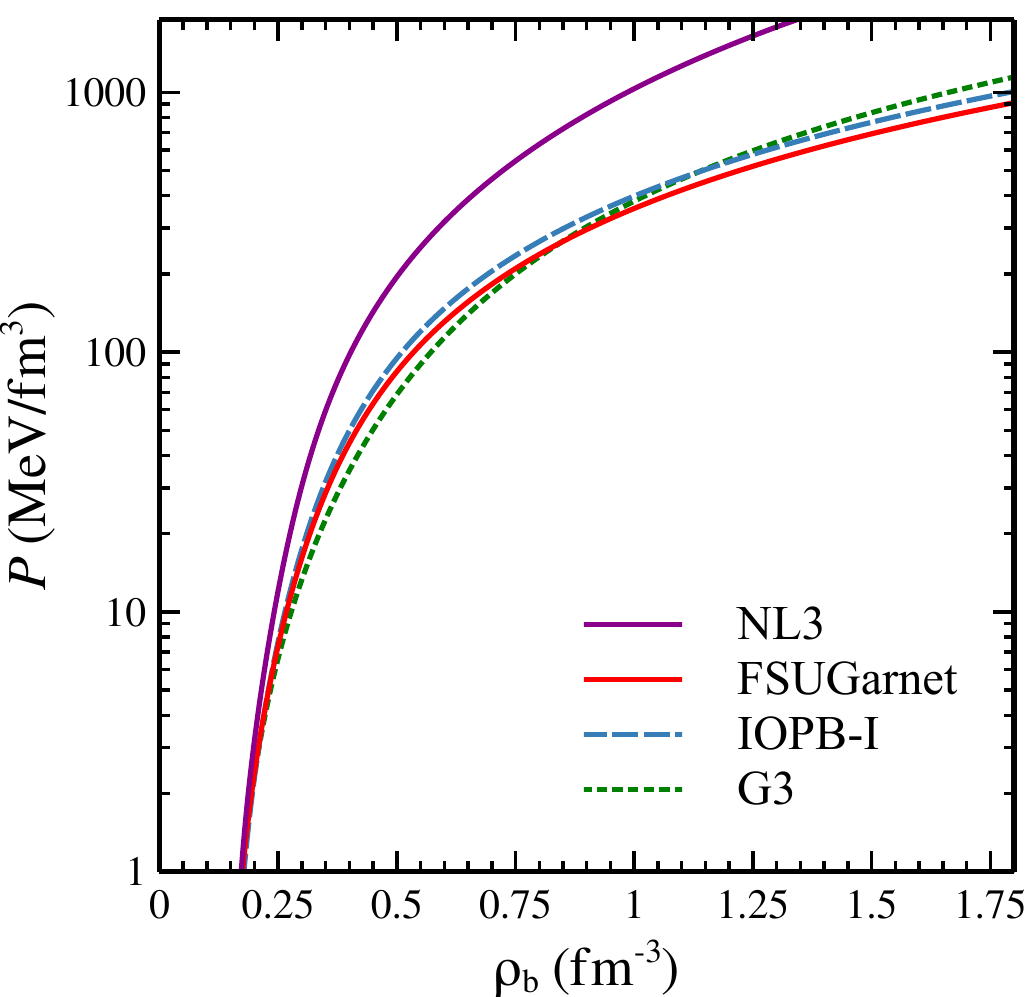}
    \caption{EOSs of the Nuclear matter for NL3 set with other three considered sets.}
    \label{fig:nm_eos}
\end{figure}

For a comparison, we plot the EoS of the nuclear matter for three considered E-RMF parameter sets along with one RMF parameter set NL3 \cite{Lalazissis_1997} in Fig. \ref{fig:nm_eos}. It is observed that the NL3 is the stiffest EoS compared to the other three E-RMF sets. Hence, the predicted NM properties such as incompressibility, symmetry energy and its slope parameter etc. for NL3 case is quite larger as compared to other three as shown in Table \ref{bulkproperties}. Also the predicted properties does not satisfy the empirical/experimental data.
On the other hand, E-RMF  parameters satisfy various constraints on EoS and are used in this work for the complete description of the neutron star. The structure and properties of the inner crust are calculated using the famous CLDM, assuming the existence of spherical clusters surrounded by the gas of dripped neutrons throughout the inner crust. The bulk energy of the cluster in Eq. (\ref{ecluster}) and neutron gas is calculated using the E-RMF parameter sets FSUGarnet, IOPB-I, and G3, ensuring numerical and physical consistency. 

\begin{figure*}
    \centering
    \includegraphics[width=0.8\textwidth]{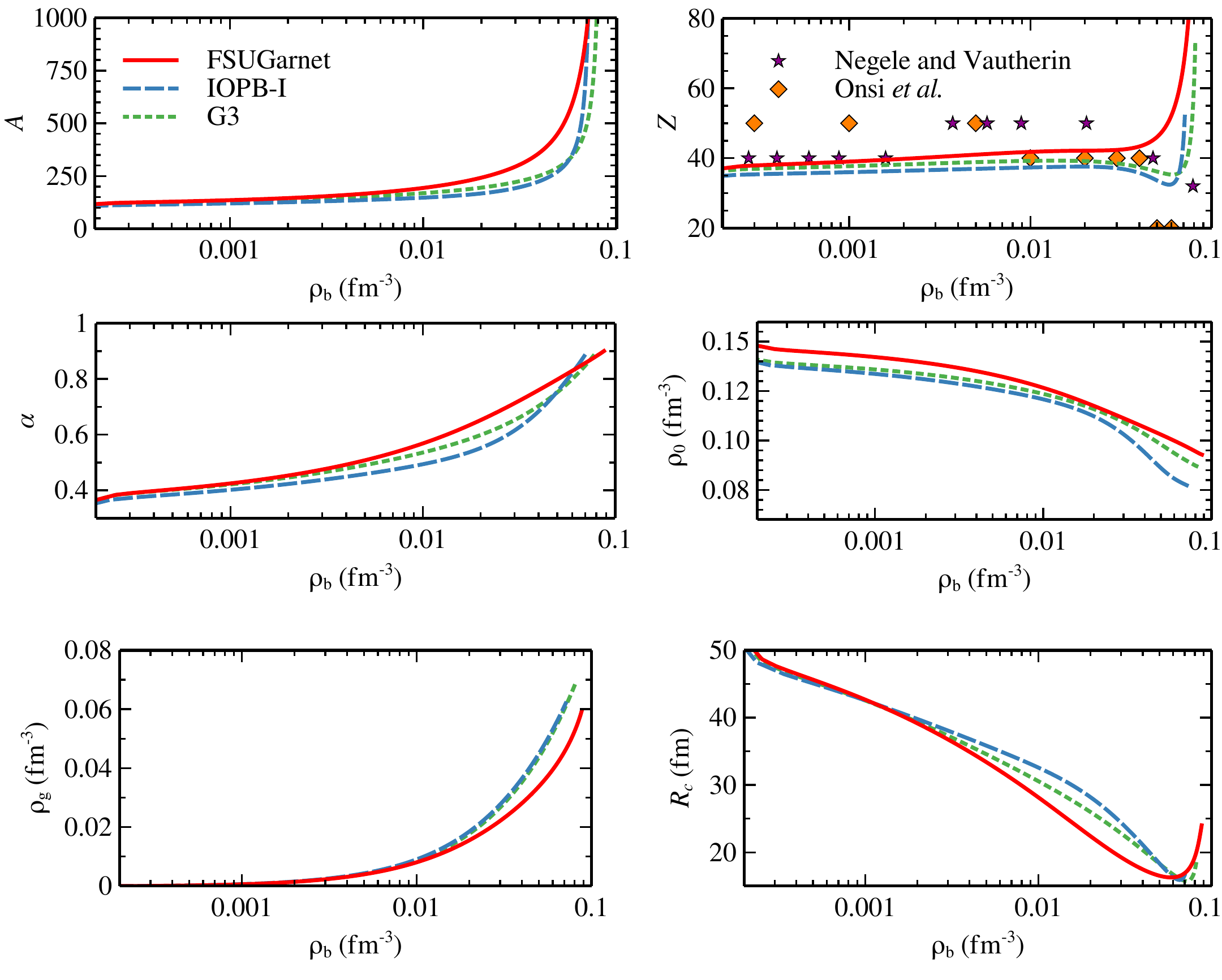}
    \caption{The variation of mass number $A$, proton number $Z$, asymmetry $\alpha$, average cluster density $\rho_0$, the neutron gas density $\rho_g$ and the radius of cell with the baryon density $\rho_b$ in the inner crust of neutron star with FSUGarnet, IOPB-I, and G3 E-RMF parameter set. The quantum calculation by Negele and Vautherin \cite{NEGELE1973298} and Onsi {\it {\it et al.}} \cite{Onsi_2008} are also shown.}
    \label{fig:icprop}
\end{figure*}

The most important aspect in the calculation of inner crust structure is the parametrization of the surface and curvature energy of the cluster. The curvature energy helps to understand the surface energy of the cluster better and is an integral part of the modified liquid-drop formulas \cite{Pomorski_2003}. Since we do not have the significant knowledge of surface energy of very neutron-rich nuclei from the laboratory experiments, we resort to the fitting of semiempirical formula such as given in Eq. (\ref{ecluster}).  In order to fit the surface and curvature energy of CLDM  with the experimental mass, we define a parameter space $\boldsymbol{S}= \{\sigma_0, b_s, \sigma_{0,c}, \beta,\alpha,p\}$ which is fitted to the experimental mass obtained from AME2020 table \cite{Huang_2021}. The goodness of reproduction of experimental binding energy is measured by the penalty function $ \chi^2(\boldsymbol{S})$ as \cite{Dobaczewski_2014}
\begin{equation}
    \chi^2(\boldsymbol{S})=\frac{1}{N}\sum_{i=1}^{N}\Big(\frac{(\mathcal{O}_i(s)-\mathcal{O}^{exp}_i)^2}{\Delta \mathcal{O}_i^2} \Big),
\end{equation}
where $N$ is the degree of freedom, $\mathcal{O}_i(s)$ stands for the calculated energy of cluster, $\mathcal{O}^{exp}_i$ for the experimental binding energy and $\Delta \mathcal{O}_i$ for adopted systematic theoretical error of 0.1 MeV \cite{carreau2020modeling}. The value of $p$, which takes care of isospin asymmetry dependence of surface energy, is taken to be 3. This is a favorable choice in  various calculations of surface energy \cite{Lattimer_1991, Avancini_2009}, and $\alpha$ is taken to be 5.5 as prescribed in \cite{Newton_2012}. The parameter space $\boldsymbol{S}$ then reduces to four variables whose values for different E-RMF parameter sets used in this study are given in Table \ref{tab:surfaceparameter}.
\begin{table}
\centering
\caption{The fitted value of surface and curvature energy parameters for the FSUGarnet, IOPB-I, and G3 force parameter set. The value of $\alpha$ and $p$ is taken to be 5.5 and 3 respectively. Experimental binding energy is taken from AME2020 table \cite{Huang_2021}. }
\label{tab:surfaceparameter}
\scalebox{1.1}{
\begin{tabular}{lllll}
\hline
\hline
Parameter & 
\begin{tabular}[c]{@{}l@{}} \hspace{0.2cm} $\sigma_0$ \\(MeV fm$^{-2}$) \end{tabular}& \hspace{0.2cm} $b_s$ &
\begin{tabular}[c]{@{}l@{}} \hspace{0.2cm} $\sigma_{0,c}$ \\(MeV fm$^{-1}$) \end{tabular} & \hspace{0.2cm} $\beta$ \\
\hline
FSUGarnet &   1.13975            & 29.39987      &   0.07819  &  0.44021\\ \hline
IOPB-I    &   0.97594            & 16.35460      &   0.09064  &  0.81485\\ \hline
G3        &   0.88424            & 26.58373      &   0.09921  &  0.93635\\ 
\hline
\hline
\end{tabular}%
}
\end{table}

The importance of fitting individual parameter set for the experimental mass excess instead of taking the same value for all the parameter sets is clear from the Table \ref{tab:surfaceparameter}, where one can see a substantial difference in fitted parameters of surface and curvature energy. The neutron star's inner crust and crustal properties are susceptible to the surface and curvature energy, making this step essential for the CLDM calculation. It is also clear from Table \ref{tab:surfaceparameter} that the fitting process underestimates the value of $\sigma_0$ and $\sigma_{0,c}$ as all other energies such as deformations are included in these parameters themselves. 

After fixing the surface parameters, we now calculate the composition of the neutron star inner crust, which is shown in Fig. \ref{fig:icprop} as a function of baryon density for the FSUGarnet, G3, and IOPB-I parameter sets. The number of nucleons $A$ inside the cluster increase  monotonically with increasing density. One can see a steep rise in the number of nucleons when approaching the crust-core transition density, thereby indicating that the matter is transiting to a homogeneous phase of nucleons and leptons.  The variation of charge number is also shown in Fig. \ref{fig:icprop}. It is observed that the $Z\approx$ 40 dominates over the majority of the inner crust. This feature is analogous with the quantum calculation carried by Negele and Vautherin \cite{NEGELE1973298} which predicts the dominance of $Z=40$ at lower densities and $Z=50$ at higher densities along with the calculations by Onsi {\it {\it et al.}} \cite{Onsi_2008}. The distinctive feature of these works is the existence of strong proton quantum-shell effects in the nuclear cluster with $Z=40$ and $50$ in the inner crust of the neutron star. One may note that the $Z=40$ is not a magic number in ordinary nuclei but corresponds to a filled proton subshell.  Recent calculation by BCPM \cite{BKS_2015} and D1M \cite{sym13091613} also indicated the same feature of inner crust. 
\begin{figure}
    \centering
    \includegraphics[width=0.45\textwidth]{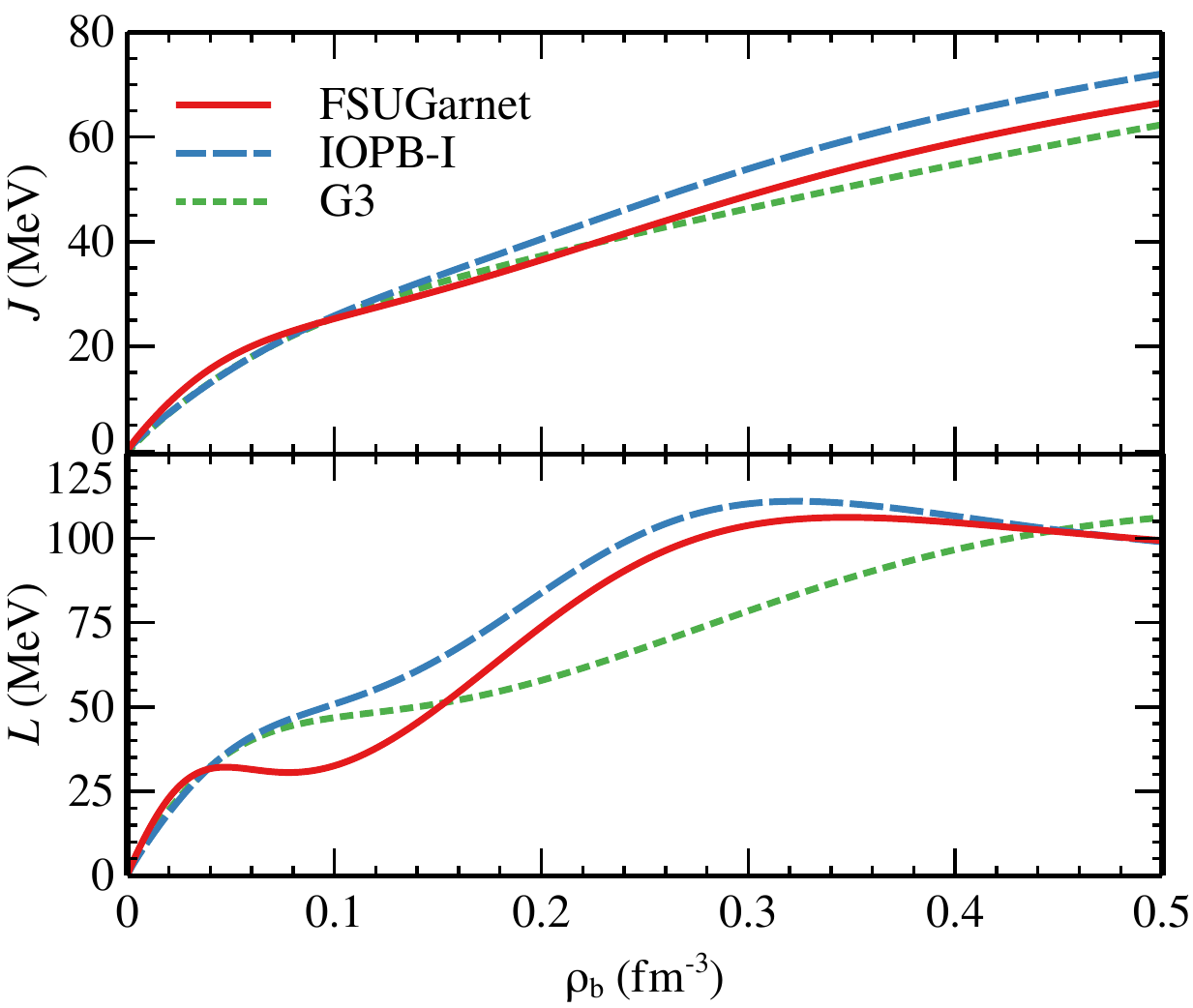}
    \caption{The density dependent symmetry energy ($J$) and slope parameter ($L$) for different E-RMF parametrizations. }
    \label{fig:symmenergy}
\end{figure}

Distribution of mass and charge number in inner crust within CLDM formalism primarily depends on two parameters; a) the isovector surface parameter $p$ in Eq. (\ref{eq:sigma}) which is responsible for the isospin dependence of surface energy, and b) the density-dependent symmetry energy or slope parameter of the EoS used to calculate the bulk energy of cluster. It is observed that the surface parameter $p=3$ correctly estimates the properties of the inner crust properties such as crust-core transition density in agreement with the dynamical \cite{Boquera_2019} or thermodynamical \cite{Bao_2020} formalisms and is used in various works such as Refs. \cite{Avancini_2009, Avancini_2008}. In the same context, we perform the inner crust calculation with $p=3$. 
Furthermore, it is an artifact of the literature that nuclear symmetry energy plays a vital role in the structural properties of a neutron star, such as radii, the moment of inertia, crust-core transition density, etc \cite{fattoyev_2013}. Additionally, it was observed in Ref. \cite{Pearson_2018} that the symmetry energy correlates with the EoS of the inner crust for the Brussels–Montreal functionals. Recently Dutra {\it {\it et al.}} \cite{dutra2021neutron}  suggested that the mass and thickness of the crust are more sensitive to the symmetry energy compared to other saturation properties. Taking motivation from these facts and to ascertain the effect of symmetry energy ($J$) and slope parameter ($L$)  on the equilibrium distribution of inner crust, we plot these quantities in Fig. \ref{fig:symmenergy} for the FSUGarnet, IOPB-I, and G3 parameter sets. All these sets follow the constraints from the experimental flow data \cite{Kumar_2018, Tsang_2012}. The behavior of $J$ and $L$ of parameter sets used is different for different density regions.  At sub-saturation densities ($<0.1$ fm$^{-3}$), which is relevant for the inner crust, the FSUGarnet shows the maximum symmetry energy followed by IOPB-I and G3. This results in the smallest slope parameter for the FSUGarnet and the highest for the G3 set. This slope parameter behavior  suggests that the higher symmetry energy or lower slope parameter of an EoS in the sub-saturation density region corresponds to the larger nucleon and charge number of clusters inside neutron star crust. This fact is also verified in Ref.   \cite{Oyamatsu_2007} which used macroscopic nuclear models to study the inner crust of the neutron star.
\begin{figure}
    \centering
    \includegraphics[width=0.48\textwidth]{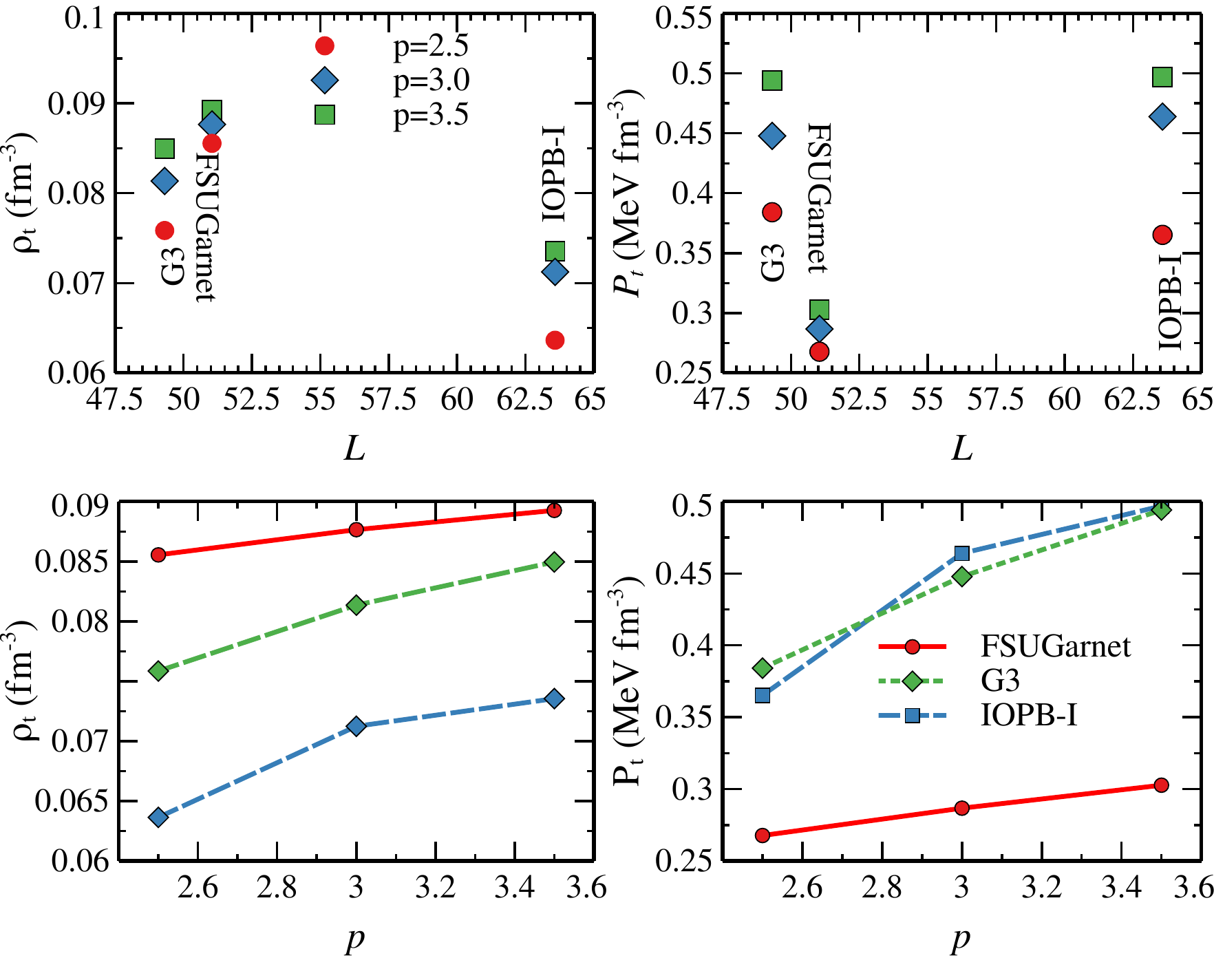}
    \caption{Crust-core transition density and pressure as a function of slope parameter $L$ and $p$ (Eq. (\ref{eq:sigma})) for the FSUGarnet, IOPB-I, and G3 parameter sets.}
    \label{fig:cctransition}
\end{figure}

With increasing density or distance from the star's surface, the spherical cluster becomes more and more asymmetric and dilute. The asymmetry $\left(\alpha=\frac{\rho_n-\rho_p}{\rho_n+\rho_p}\right)$ reaches $\approx 0.9$ when reaching the crust boundary, and the density of cluster ($\rho_0$) becomes comparable to the density of neutron gas ($\rho_g$) surrounding these clusters. It should be mentioned that the terms associated with iso-vector meson coupling affect the asymmetricity of the system. But in accordance to the mathematical conventions, the terms with high powers of iso-vector mesons are less effective, so, the linear term decides the asymmetry factor considerably. We checked the mentioned asymmetry value for other usual RMF models too and did not observe any major change for the same. The  asymmetry at crust boundary are 0.896, 0.900, 0.902, \& 0.894 for NL3, FSUGarnet, IOPB-I and G3 sets respectively. However, the FSUGarnet shows the largest asymmetry and density of cluster as one starts moving toward the core from the outer crust of neutron star, while IOPB-I the least owing to the behavior of their symmetry energy. Finally, the radius of the WS cell decreases with density while the cluster keeps growing in size. This leads the cluster to get closer and closer to form a large cluster and ultimately convert to homogeneous matter when reaching the crust-core boundary. The slope parameter has an inverse effect on the density of neutron gas and WS cell radius. A larger $L$ corresponds to the smaller neutron gas density and radius of the cluster. 

We study the crust-core transition from the crust side shown in Fig. \ref{fig:cctransition} using Eq. (\ref{eq:cctransition}). As discussed, the EoS of the inner crust is sensitive to the choice of surface parameters $p$ and $L$. To investigate this, we plot the transition density $\rho_t$ and pressure $P_t$ as a function of $L$ and $p$. The G3 parameter set predicts a larger transition density as compared to the IOPB-I set owing to its smaller $L$, while FSUGarnet does not follow the trend. In general practice, the crust-core transition density and pressure are anti-correlated to the saturation value of $L$ for a given EoS. However, one can notice in Fig. \ref{fig:symmenergy} that the behavior of $L$ is different for below and above saturation density. Therefore, if we consider the behavior of $L$ in the subsaturation density region, the trends in the crust-core transition density could be understood more precisely. The FSUGarnet set with the least $L$ estimates the larges transition density, and IOPB-I with maximum $L$ estimates the lowest crust-core transition density. The transition pressure follows the same trend, however, in the opposite way. The isovector surface parameter $p$ seems to act similarly to the symmetry energy. The transition pressure and density are positively correlated with the value of $p$. This fact suggests the importance of isospin-dependent surface tension in the CLDM calculation of inner crust. Furthermore, the correlation of transition density and pressure of crust-core transition is in harmony with the trends obtained from \cite{Lattimer_2007}. Recently Bao-An Li and Macon Magno \cite{Bao_2020} found that the curvature $K_{sym}$ plays a more important role than the slope $L$ in determining the crust-core transition density using the EoSs generated from meta-modeling. We also find a similar behavior of $\rho_t$ while comparing the value of $K_{sym}$ from Table \ref{bulkproperties}. 
\begin{figure}
    \centering
    \includegraphics[width=0.45\textwidth]{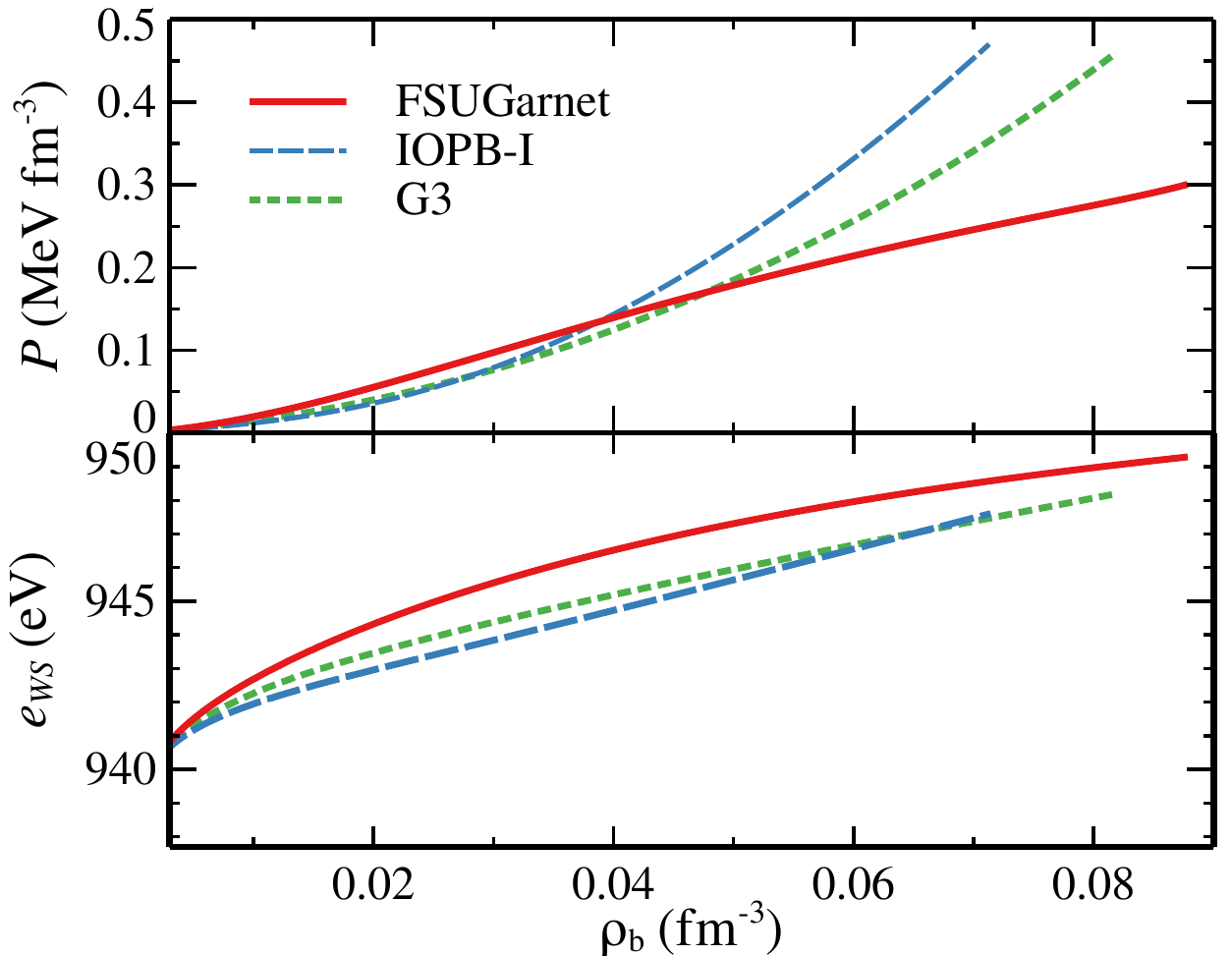}
    \caption{The EoS for the inner crust and equilibrium value of WS cell energy using the E-RMF parameter sets FSUGarnet, IOPB-I and G3. }
    \label{fig:eosic}
\end{figure}

It is clear from the above discussion that the structure of the inner crust is susceptible to the behavior of density-dependent symmetry energy and slope parameter in the sub-saturation density region. In the E-RMF framework, the symmetry energy is controlled mainly by the cross-coupling ($\Lambda_\omega$) of isoscalar-vector ($\omega$) and isovector-vector ($\rho$) mesons [see Eq. (\ref{rmftlagrangian})]. In addition, the parameter set  G3  takes the $\delta$ meson as the additional degree of freedom which helps to change the variation of $L$ and $J$ to reproduce the theoretical and observational constraints \cite{Singh_2014}. The $J$ and $L$ also play a crucial role in estimating the instability in the homogeneous nuclear matter \cite{Vishal_2021}. Therefore, $\Lambda_\omega$ becomes an essential parameter in the E-RMF forces that govern various aspects of the neutron star structure. 

In Fig. \ref{fig:eosic} we show the EoS of the inner crust for the FSUGarnet, IOPB-I, and G3 E-RMF parameter sets along with the WS cell energy and the tabulated data in Table \ref{tab:iceosdata}. One may see that the inner crust is primarily  model-dependent, where the stiffness is related to the behavior of symmetry energy or slope parameter. Higher symmetry energy at subsaturation densities corresponds to the larger $e_{WS}$, which is the case with FSUGarnet in Fig. \ref{fig:eosic}.  The behavior of G3 and IOPB-I is similar, with IOPB-I estimating a comparatively stiffer EoS which is also in accordance with the behavior of the symmetry energy. Therefore, we believe that the symmetry energy and its derivative predominantly decide the inner crust structure. However, one needs a detailed statistical study of various E-RMF parameter sets (e.g., Bayesian and correlation analysis) to comment on the ambiguities.
One may further note that, unlike in the outer crust, the pressure of the inner crust is mainly dependent on the neutron gas surrounding the clusters. Therefore, the parameters used must follow the necessary constraints on the pure neutron matter (PNM). It is seen that the FSUGarnet, IOPB-I, and G3 reasonably satisfy the results obtained using microscopic chiral EFT \cite{Vishal_2021}, making these parameters suitable for the calculation of inner crust EoS. 

\begin{figure}
    \centering
    \includegraphics[width=0.48\textwidth]{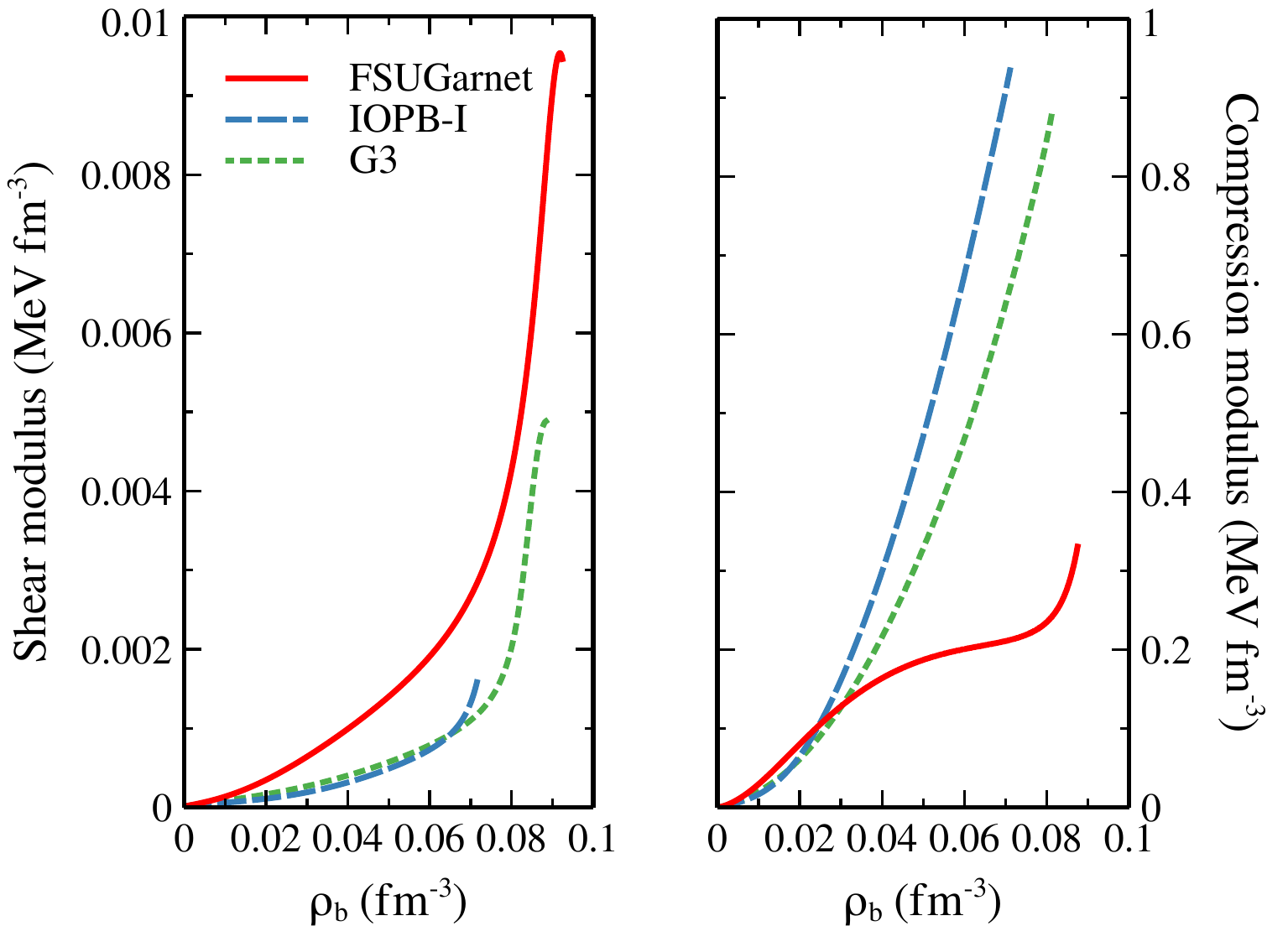}
    \caption{The effective shear and compression modulus for BCC lattice in the inner crust of neutron stare using the FSUGarnet, IOPB-I, and G3 parameter sets.}
    \label{fig:modulus}
\end{figure}

It should be noted that in this work, we restrict ourselves to  spherically symmetric WS cell for the calculation of  inner crust of the neutron star. However, as one approaches the crust-core boundary,  there might be an energetic preference for nonspherical shapes (rod, slab, tube, bubble, etc.) commonly known as ``nuclear pasta" \cite{Ravenhall_1983, Avancini_2008, Martin_2015, Maruyama_1998}. These structures influence various properties of neutron star crust such as crustal oscillation modes, crust cooling, crust shattering, magnetic field evolution, etc \cite{Newton_2021}. Nevertheless, it is  seen that the existence of pasta structure is sensitive to the approximations made and  minute energy differences exist between spherical and nonspherical cell shapes. Therefore, nuclear pasta structures have a weak impact on the EoS  \cite{Pearson_2018} and the WS cell composition \cite{Pearson_2020}  and hence they do not affect the global properties of neutron stars, such as the mass-radius profile. However,  for the quantitative analysis of pasta structure, we shall carry a comprehensive study of neutron star crust including all possible structures  in a forthcoming assignment. 

It is shown that the fundamental seismic shear mode, observed as a quasiperiodic oscillation in giant flares emitted by highly magnetized neutron stars, is particularly sensitive to the EoS of crust \cite{Steiner_2009, Sotani_2012}. In that context, we assume the neutron star crust as an isotropic BCC poly-crystal whose elastic properties are a function of two elastic moduli: shear ( $\mu$) and compression modulus ($K$). These are written as \cite{Haensel_2008}
\begin{equation}
    \begin{aligned}
    &K=\rho_b\frac{\partial P}{\partial \rho_b}=\Gamma P,\\
    &\mu=0.1194\frac{\rho_i (Ze)^2}{R_{cell}},
    \end{aligned}
\end{equation}
where $\Gamma$ is the adiabatic index and $\rho_i$ is the density of
nuclei. The variation of shear and compression modulus as a function of baryon density is shown in Fig. \ref{fig:modulus}. The shear modulus depends on the distribution of $Z$ and the size of the cell, which is a smoothly increasing function of average baryon density as shown in Fig. \ref{fig:icprop}. As a result, the shear modulus increases continuously on moving toward the core. The FSUGarnet and IOPB-I show the maximum and minimum values of $\mu$. A higher value of $\mu$ means that the fundamental shear mode will have a higher frequency. The compression modulus also increases with density and has an opposite trend as compared to the shear modulus. 
\begin{figure}
    \centering
    \includegraphics[width=0.45\textwidth]{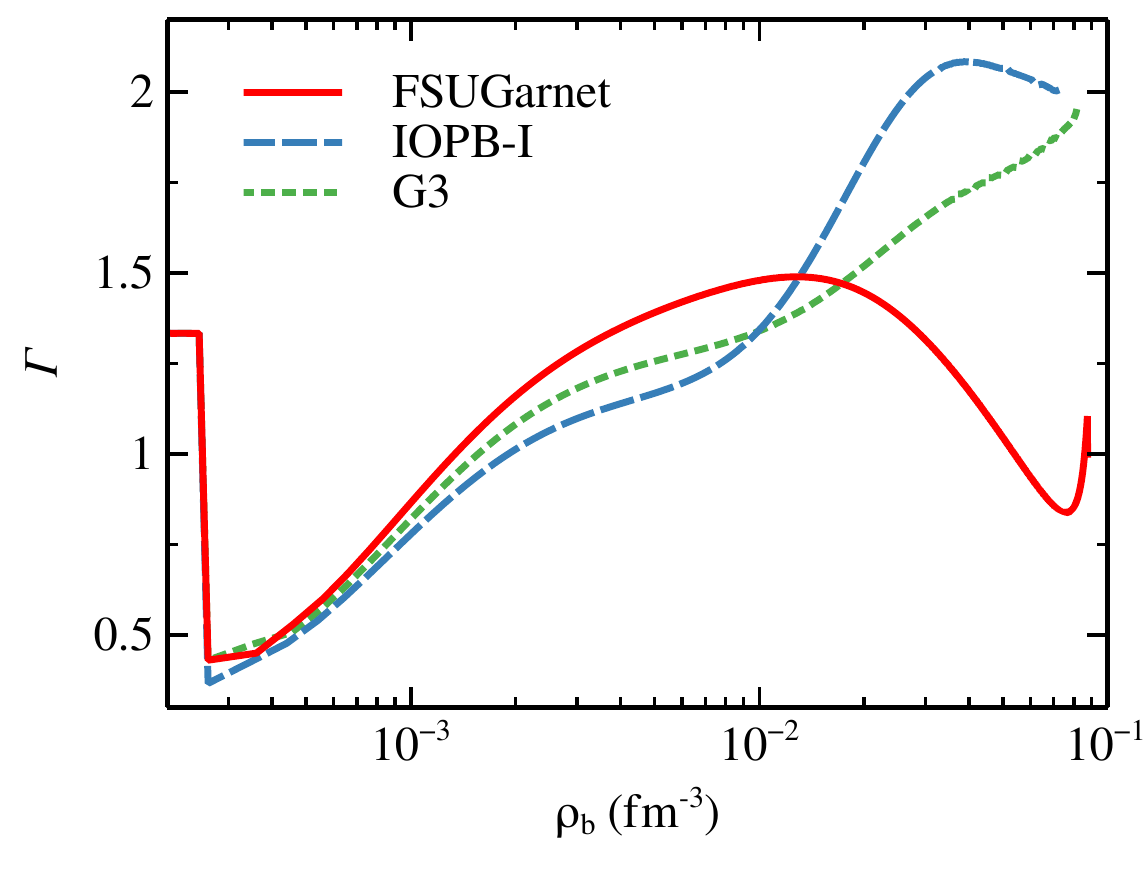}
    \caption{Adiabatic index of the inner crust calculated from the FSUGarnet, IOPB-I, and G3 E-RMF forces.}
    \label{fig:adiabatic}
\end{figure}

Finally, the adiabatic index, which determines the response of the crust toward the compression and decompression, is plotted in Fig. \ref{fig:adiabatic} from the outer layer of outer crust till the transition of inner crust to the core. As the pressure in the outer crust is prominently determined from the ultrarelativistic electron gas,  the $\Gamma$ becomes equal to 4/3. The onset of the inner crust is marked by dripped neutrons which soften the EoS. This results in a decrease in the value of $\Gamma$ considerably. As the density in the crust increases, the neutron gas density increases resulting in more and more pressure of neutron gas. As a consequence, the $\Gamma$ increases and reaches up to $\approx$ 2 on reaching the crust-core transition. The FSUGarnet shows a relatively lower value of $\Gamma$ at CC point, which can be explained based on the behaviors of its compression modulus in Fig. \ref{fig:modulus}. The results are in agreement with the microscopic calculation using  three-body forces\cite{BKS_2015}.
\begin{table*}
\centering
\caption{Composition and EoS of inner crust with the IOPB-I, FSUGarnet, and G3 E-RMF parameter sets. The table include the values of pressure ($P$), energy (${\cal E}$), mass ($A$) and charge ($Z$) of the cluster and the radius ($R_c$) of cell.}
\label{tab:iceosdata}
\renewcommand{\arraystretch}{1.6}
\resizebox{\textwidth}{!}{%
\begin{tabular}{l|lllll|lllll|lllll}
\hline
\hline
&\multicolumn{5}{c|}{IOPB-I} & \multicolumn{5}{c|}{FSUGarnet} & \multicolumn{5}{c}{G3} \\ 
\hline
\cmidrule(lr){2-6}\cmidrule(lr){7-11}\cmidrule(lr){12-16}
\begin{tabular}[c]{@{}l@{}} \hspace{0.2cm} $\rho_b$ \\(fm$^{-3}$) \end{tabular}&
\begin{tabular}[c]{@{}l@{}} \hspace{0.5cm} $P$ \\(MeV fm$^{-3}$) \end{tabular} &
\begin{tabular}[c]{@{}l@{}} \hspace{0.5cm} ${\cal E}$ \\(MeV fm$^{-3}$) \end{tabular}&
\hspace{0.2cm} $A$ &
\hspace{0.2cm} $Z$ & 
\begin{tabular}[c]{@{}l@{}} \hfill $R_c$ \\(fm) \end{tabular}&
\begin{tabular}[c]{@{}l@{}} \hspace{0.5cm} $P$ \\(MeV fm$^{-3}$) \end{tabular} &
\begin{tabular}[c]{@{}l@{}} \hspace{0.5cm} ${\cal E}$ \\(MeV fm$^{-3}$) \end{tabular}&
\hspace{0.2cm} $A$ &
\hspace{0.2cm} $Z$ & 
\begin{tabular}[c]{@{}l@{}} \hfill $R_c$ \\(fm) \end{tabular}&
\begin{tabular}[c]{@{}l@{}} \hspace{0.5cm} $P$ \\(MeV fm$^{-3}$) \end{tabular} &
\begin{tabular}[c]{@{}l@{}} \hspace{0.5cm} ${\cal E}$ \\(MeV fm$^{-3}$) \end{tabular}&
\hspace{0.2cm} $A$ &
\hspace{0.2cm} $Z$ & 
\begin{tabular}[c]{@{}l@{}} \hfill $R_c$ \\(fm) \end{tabular}
\\
\hline
0.0003 & 0.000512 & 0.281414 & 112.7301 & 35.3635 & 47.0669 & 0.000529 & 0.281392 & 124.2469 & 37.9120 & 47.7178 & 0.000529 & 0.281396 & 121.0933 & 36.9770 & 47.3051 \\
0.0023 & 0.002196 & 2.162921 & 127.4424 & 36.4676 & 39.2188 & 0.002666 & 2.163182 & 148.1375 & 39.9832 & 38.3637 & 0.002421 & 2.163018 & 139.7759 & 38.3019 & 38.6883 \\
0.0043 & 0.004378 & 4.046355 & 134.4834 & 36.8530 & 36.5161 & 0.006011 & 4.047617 & 162.3130 & 40.8027 & 34.4444 & 0.005088 & 4.046883 & 149.7114 & 38.7594 & 35.4098 \\
0.0063 & 0.006857 & 5.930916 & 139.6914 & 37.0827 & 34.8056 & 0.010280 & 5.933861 & 174.1880 & 41.2951 & 31.7416 & 0.008233 & 5.932151 & 157.3900 & 39.0083 & 33.2786 \\
0.0083 & 0.009637 & 7.816308 & 144.0866 & 37.2410 & 33.5129 & 0.015351 & 7.821608 & 184.9821 & 41.6146 & 29.6417 & 0.011770 & 7.818483 & 163.9397 & 39.1544 & 31.6794 \\
0.0103 & 0.012790 & 9.702446 & 148.1121 & 37.3580 & 32.4303 & 0.021143 & 9.710712 & 195.1915 & 41.8254 & 27.9171 & 0.015679 & 9.705767 & 169.8407 & 39.2376 & 30.3825 \\
0.0123 & 0.016404 & 11.589212 & 152.0036 & 37.4464 & 31.4627 & 0.027576 & 11.600992 & 205.0751 & 41.9625 & 26.4566 & 0.019966 & 11.593800 & 175.3510 & 39.2765 & 29.2779 \\
0.0143 & 0.020568 & 13.476612 & 155.9021 & 37.5109 & 30.5624 & 0.034569 & 13.492286 & 214.7894 & 42.0488 & 25.1970 & 0.024647 & 13.482594 & 180.6262 & 39.2813 & 28.3051 \\
0.0163 & 0.025362 & 15.364671 & 159.8925 & 37.5534 & 29.7039 & 0.042041 & 15.384688 & 224.4444 & 42.1009 & 24.0977 & 0.029740 & 15.372041 & 185.7704 & 39.2576 & 27.4281 \\
0.0183 & 0.030855 & 17.253335 & 164.0333 & 37.5730 & 28.8741 & 0.049912 & 17.277980 & 234.1273 & 42.1316 & 23.1307 & 0.035264 & 17.262138 & 190.8534 & 39.2087 & 26.6240 \\
0.0203 & 0.037103 & 19.142629 & 168.3629 & 37.5681 & 28.0663 & 0.058107 & 19.172182 & 243.9109 & 42.1518 & 22.2749 & 0.041231 & 19.152846 & 195.9276 & 39.1367 & 25.8779 \\
0.0223 & 0.044151 & 21.032522 & 172.9082 & 37.5368 & 27.2774 & 0.066553 & 21.067174 & 253.8634 & 42.1702 & 21.5146 & 0.047654 & 21.044189 & 201.0313 & 39.0433 & 25.1792 \\
0.0243 & 0.052030 & 22.923195 & 177.6925 & 37.4765 & 26.5066 & 0.075186 & 22.962877 & 264.0506 & 42.1949 & 20.8366 & 0.054541 & 22.936130 & 206.1971 & 38.9293 & 24.5210 \\
0.0263 & 0.060762 & 24.814495 & 182.7338 & 37.3848 & 25.7538 & 0.083950 & 24.859370 & 274.5427 & 42.2329 & 20.2305 & 0.061897 & 24.828700 & 211.4524 & 38.7959 & 23.8976 \\
0.0283 & 0.070358 & 26.706495 & 188.0512 & 37.2593 & 25.0194 & 0.092795 & 26.756579 & 285.4087 & 42.2906 & 19.6879 & 0.069724 & 26.721787 & 216.8215 & 38.6439 & 23.3053 \\
0.0303 & 0.080822 & 28.599194 & 193.6626 & 37.0978 & 24.3043 & 0.101674 & 28.654369 & 296.7274 & 42.3745 & 19.2015 & 0.078025 & 28.615431 & 222.3302 & 38.4740 & 22.7407 \\
0.0323 & 0.092152 & 30.492652 & 199.5922 & 36.8984 & 23.6087 & 0.110554 & 30.552771 & 308.5798 & 42.4907 & 18.7654 & 0.086799 & 30.509675 & 228.0027 & 38.2873 & 22.2015 \\
0.0343 & 0.104339 & 32.386812 & 205.8659 & 36.6596 & 22.9332 & 0.119398 & 32.451663 & 321.0590 & 42.6453 & 18.3746 & 0.096044 & 32.404474 & 233.8671 & 38.0848 & 21.6857 \\
0.0363 & 0.117373 & 34.281724 & 212.5165 & 36.3809 & 22.2781 & 0.128181 & 34.351065 & 334.2689 & 42.8449 & 18.0250 & 0.105759 & 34.299832 & 239.9538 & 37.8677 & 21.1916 \\
0.0383 & 0.131239 & 36.177311 & 219.5887 & 36.0623 & 21.6434 & 0.136879 & 36.250965 & 348.3255 & 43.0963 & 17.7132 & 0.115943 & 36.195666 & 246.2988 & 37.6376 & 20.7177 \\
0.0403 & 0.145923 & 38.073709 & 227.1356 & 35.7053 & 21.0292 & 0.145474 & 38.151365 & 363.3607 & 43.4070 & 17.4362 & 0.126592 & 38.092189 & 252.9423 & 37.3964 & 20.2631 \\
0.0423 & 0.161410 & 39.970881 & 235.2303 & 35.3131 & 20.4360 & 0.153951 & 40.052167 & 379.5305 & 43.7854 & 17.1918 & 0.137707 & 39.989108 & 259.9348 & 37.1462 & 19.8267 \\
0.0443 & 0.177688 & 41.868732 & 243.9690 & 34.8913 & 19.8641 & 0.162296 & 41.953359 & 397.0108 & 44.2410 & 16.9782 & 0.149287 & 41.886623 & 267.3362 & 36.8900 & 19.4077 \\
0.0463 & 0.194741 & 43.767368 & 253.4835 & 34.4481 & 19.3141 & 0.170499 & 43.854858 & 416.0135 & 44.7846 & 16.7940 & 0.161332 & 43.784718 & 275.2192 & 36.6310 & 19.0056 \\
0.0483 & 0.212560 & 45.666732 & 263.9520 & 33.9958 & 18.7874 & 0.178552 & 45.756858 & 436.7864 & 45.4290 & 16.6382 & 0.173844 & 45.683268 & 283.6740 & 36.3737 & 18.6201 \\
0.0503 & 0.231136 & 47.566890 & 275.6233 & 33.5507 & 18.2858 & 0.186450 & 47.659061 & 459.6274 & 46.1891 & 16.5101 & 0.186825 & 47.582352 & 292.8160 & 36.1233 & 18.2510 \\
0.0523 & 0.250460 & 49.467791 & 288.8484 & 33.1352 & 17.8120 & 0.194192 & 49.561660 & 484.8955 & 47.0827 & 16.4096 & 0.200280 & 49.482021 & 302.7876 & 35.8868 & 17.8984 \\
0.0543 & 0.270527 & 51.369389 & 304.1277 & 32.7789 & 17.3698 & 0.201772 & 51.464453 & 513.0301 & 48.1317 & 16.3370 & 0.214213 & 51.382168 & 313.7734 & 35.6728 & 17.5629 \\
0.0563 & 0.291330 & 53.271847 & 322.1954 & 32.5213 & 16.9645 & 0.209193 & 53.367562 & 544.5713 & 49.3622 & 16.2927 & 0.228631 & 53.282866 & 326.0136 & 35.4924 & 17.2451 \\
0.0583 & 0.312864 & 55.174940 & 344.1495 & 32.4164 & 16.6034 & 0.216456 & 55.270952 & 580.1935 & 50.8072 & 16.2781 & 0.243539 & 55.184066 & 339.8258 & 35.3602 & 16.9466 \\
0.0603 & 0.335122 & 57.078838 & 371.7034 & 32.5429 & 16.2972 & 0.223564 & 57.174654 & 620.7521 & 52.5076 & 16.2947 & 0.258946 & 57.085810 & 355.6361 & 35.2952 & 16.6694 \\
0.0623 & 0.358096 & 58.983547 & 407.6620 & 33.0209 & 16.0623 & 0.230523 & 59.078454 & 667.3394 & 54.5159 & 16.3450 & 0.274861 & 58.988009 & 374.0316 & 35.3235 & 16.4164 \\
0.0643 & 0.381778 & 60.888911 & 456.9247 & 34.0513 & 15.9241 & 0.237340 & 60.982554 & 721.3793 & 56.8996 & 16.4321 & 0.291295 & 60.890753 & 395.8363 & 35.4807 & 16.1918 \\
0.0663 & 0.406154 & 62.795116 & 528.8127 & 36.0048 & 15.9275 & 0.244023 & 62.886854 & 784.7565 & 59.7475 & 16.5605 & 0.308257 & 62.794055 & 422.2481 & 35.8176 & 16.0017 \\
0.0683 & 0.431207 & 64.701956 & 643.1830 & 39.6507 & 16.1563 & 0.250586 & 64.791356 & 860.0130 & 63.1784 & 16.7356 & 0.325756 & 64.697854 & 455.0602 & 36.4091 & 15.8547 \\
0.0703 & 0.456922 & 66.609668 & 848.7924 & 46.8294 & 16.7832 & 0.257041 & 66.696056 & 950.6534 & 67.3544 & 16.9651 & 0.343803 & 66.602142 & 497.0797 & 37.3702 & 15.7640 \\
0.0723 &  &  &  &  &  & 0.263407 & 68.600956 & 1061.6230 & 72.5019 & 17.2589 & 0.362407 & 68.506998 & 552.9294 & 38.8868 & 15.7498 \\
0.0743 &  &  &  &  &  & 0.269708 & 70.505948 & 1200.0577 & 78.9449 & 17.6307 & 0.381575 & 70.412409 & 630.7147 & 41.2793 & 15.8444 \\
0.0763 &  &  &  &  &  & 0.275974 & 72.411149 & 1376.5995 & 87.1585 & 18.0986 & 0.401315 & 72.318190 & 745.8048 & 45.1431 & 16.1026 \\
0.0783 &  &  &  &  &  & 0.282245 & 74.316548 & 1607.4969 & 97.8569 & 18.6875 & 0.421639 & 74.224643 & 930.1836 & 51.6896 & 16.6213 \\
0.0803 &  &  &  &  &  & 0.288574 & 76.222148 & 1918.1697 & 112.1331 & 19.4311 & 0.442582 & 76.131600 & 1256.6567 & 63.5913 & 17.5768 \\
0.0823 &  &  &  &  &  & 0.295044 & 78.127848 & 2348.6311 & 131.6563 & 20.3718 &  &  &  &  &  \\
0.0843 &  &  &  &  &  & 0.301780 & 80.033651 & 2959.3448 & 158.8194 & 21.5538 &  &  &  &  &  \\
0.0863 &  &  &  &  &  & 0.309003 & 81.939748 & 3824.6786 & 196.1983 & 22.9867 &  &  &  &  &  \\
0.0883 &  &  &  &  &  & 0.316890 & 84.005883 & 4964.3837 & 242.5463 & 24.6032 &  &  &  &  & \\
\hline
\hline
\end{tabular}%
}
\end{table*}
\subsection{Neutron star unified EOS, $M-R$ relation}
\label{res:NS_EOS}
\begin{figure}
    \centering
    \includegraphics[width=0.45\textwidth]{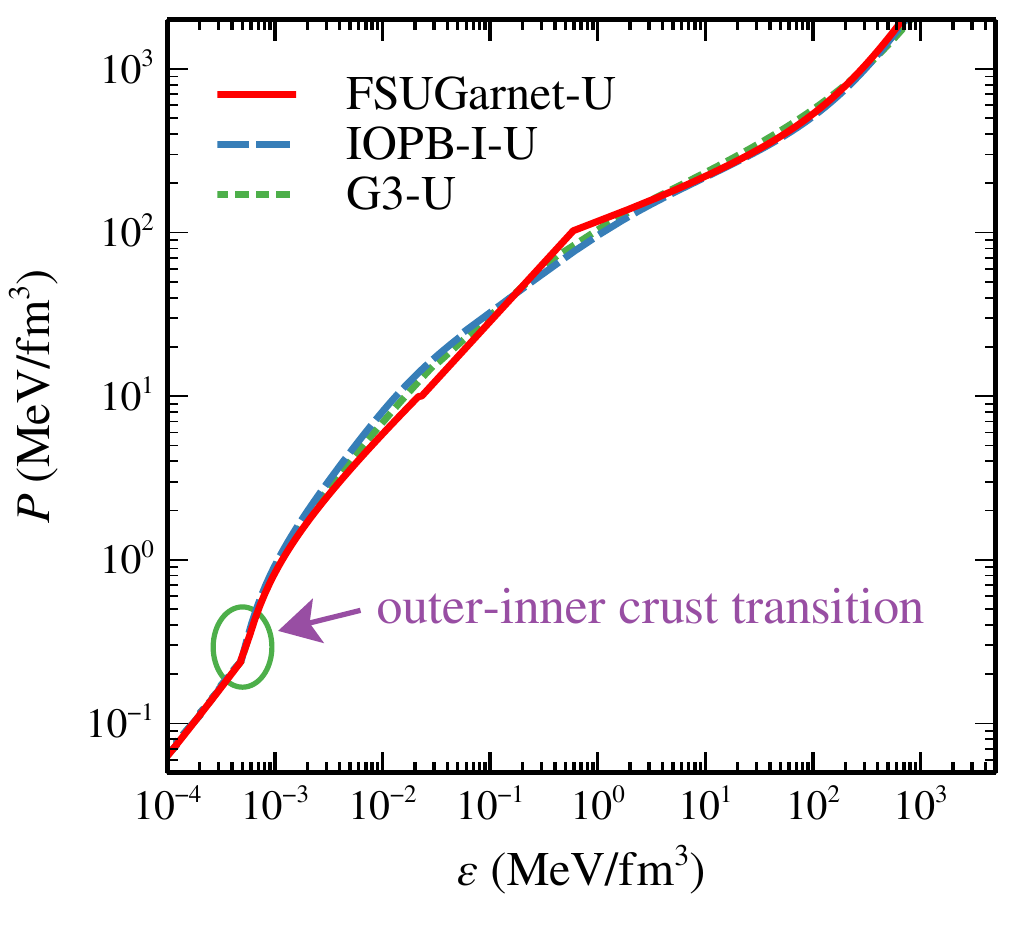}
    \caption{The unified EoSs for FSUGarnet-U, IOPB-I-U, and G3-U sets. The green line represents the outer-inner crust transition.}
    \label{fig:unifiedeos}
\end{figure}

The core EoS of the neutron star is calculated with E-RMF formalism for FSUGarnet, IOPB-I, and G3 parameter sets. For the crust part, we use both outer and inner EoS as discussed in Sec. \ref{formulaion} above. We make the unified EoS by matching the crust-core density and pressure, and is shown in Fig. \ref{fig:unifiedeos} for FSUGarnet, IOPB-I, and G3 sets.  The unified EoSs are named as FSUGarnet-U, IOPB-I-U, and G3-U, respectively and one can find from the GitHub link\footnote{{\bf \url{https://github.com/hcdas/Unified_eos}}}. The green circle represents the outer-inner crust transition. The crust-core transition is different for different forces because it is model-dependent. With these EoSs, we calculate the neutron star's mass, radius, and moment of inertia. 

\begin{figure}
\centering
\includegraphics[width=0.45\textwidth]{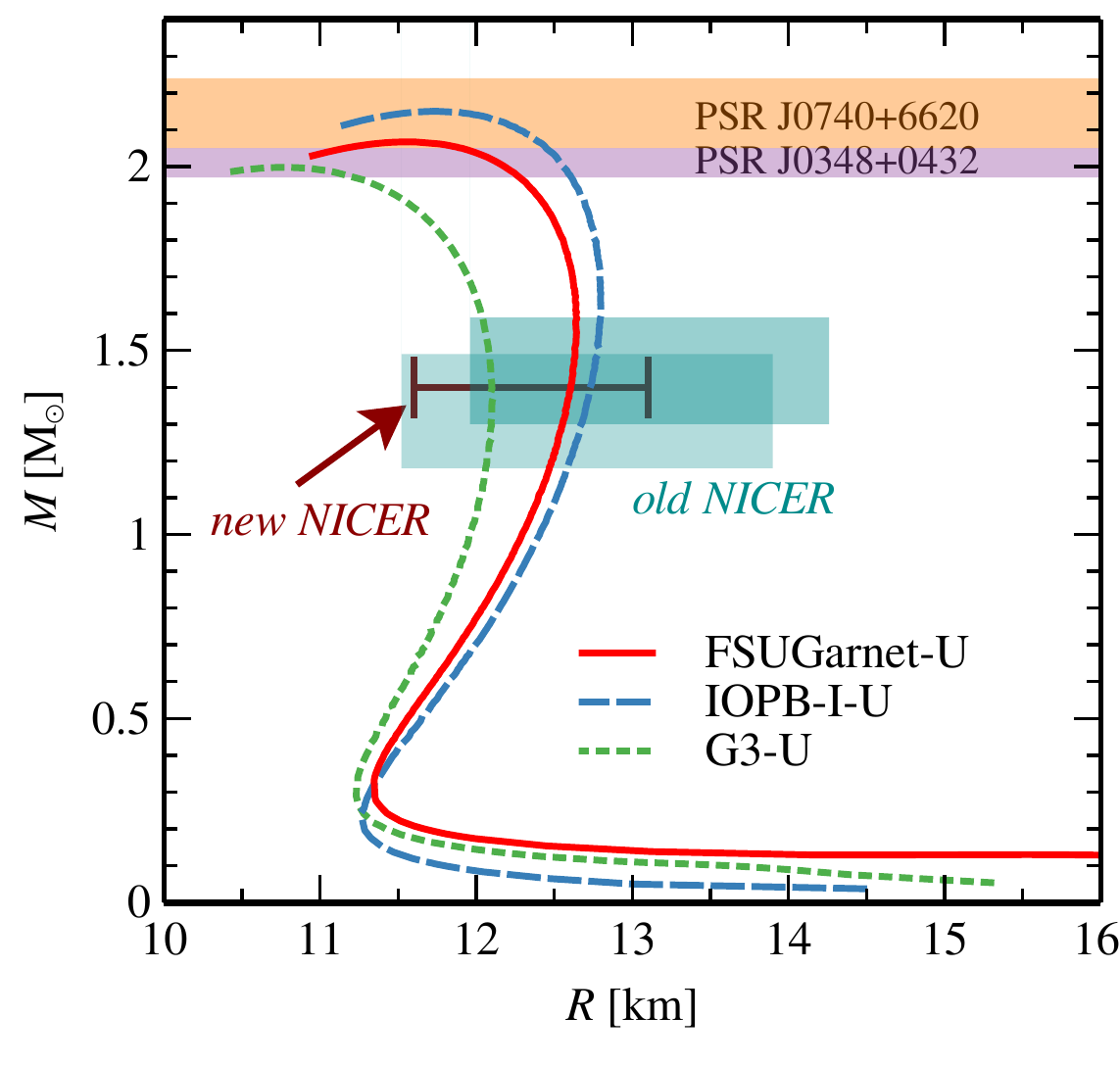}
\caption{The $M-R$ relations for three unified EoSs such as FSUGarnet-U, IOPB-I-U, and G3-U. The horizontal bars represents the PSR J0740+6620 \cite{Cromartie_2019} (light orange) and PSR J0348+0432 \cite{Antoniadis_2013} (light violet). The old NICER data are also shown with two boxes from two different analysis \cite{Miller_2019,Riley_2019}. The double-headed red line represents the radius constraints by the Miller {\it {\it {\it et al.}}} \cite{Miller_2021} for 1.4 $M_\odot$ neutron star termed as new NICER data.}
\label{fig:mr}
\end{figure}

We calculate the mass and radius of the neutron star using Eqs. (\ref{eq:pr} and \ref{eq:mr}) for a fixed central density. The $M-R$ profile is calculated for the whole star which is depicted in Fig. \ref{fig:mr} for considered sets. The maximum mass of the all the sets satisfy $\sim 2\ M_\odot$ limit. The maximum mass constraints from different massive pulsars such as PSR J0348+0432 ($M = 2.01\pm{0.04} \ M_\odot$) \cite{Antoniadis_2013} and PSR J0740+6620 ($M = 2.14_{-0.09}^{+0.10} \ M_\odot$) \cite{Cromartie_2019} are shown. The radius constraints given by Miller {\it {\it {\it et al.}}} \cite{Miller_2019} and Riley {\it {\it {\it et al.}}} \cite{Riley_2019} are shown with two dark cyan boxes termed as {\it old NICER}. The {\it new NICER} data is also shown from the study of PSR J0030+0451 with X-ray Multi-Mirror Newton for canonical star with $R_{1.4} = 12.35 \pm 0.75$ km \cite{Miller_2021}. From the figure it is clear that all the considered EoSs satisfy all constraints; such as maximum mass by two different pulsars and canonical radius by both NICER data.
\begin{figure}
    \centering
    \includegraphics[width=0.45\textwidth]{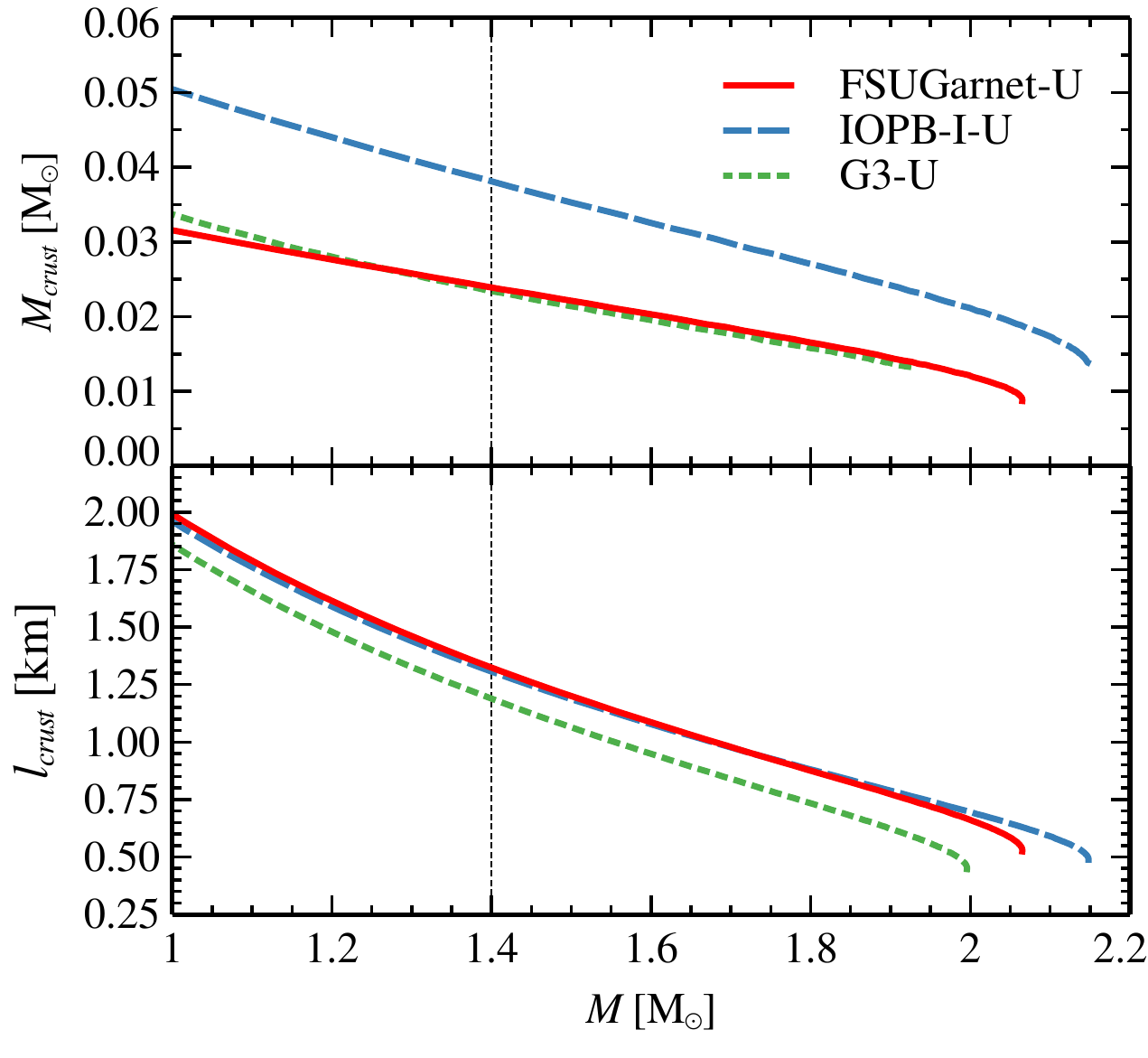}
    \caption{{\it Upper:} The mass of the crust as a function of mass for three unified EoSs. {\it Lower:} The length of the crust as a function of mass. The black dotted line represents the canonical neutron star mass.}
    \label{fig:crust}
\end{figure}

We calculate the mass and thickness of the crust for three unified EoSs using the formula $M_{crust} = M-M_{core}$, and $l_{crust} = R-R_{core}$ respectively.  The $M_{core} (R_{core})$ is the mass (radius) of the neutron star core. The variation of mass and thickness of the crust is plotted in Fig. \ref{fig:crust} for three EoSs. We ﬁnd that the crust is thicker for low mass neutron star, and it drops continuously with increasing neutron star mass. Similar results are obtained  for the crust mass as well. The  mass and thickness of the crust for all considered EoSs are given in Table \ref{tab:NS_observables}.
\subsection{Moment of inertia of the neutron star}
\label{res:MOI}
\begin{figure}
    \centering
    \includegraphics[width=0.45\textwidth]{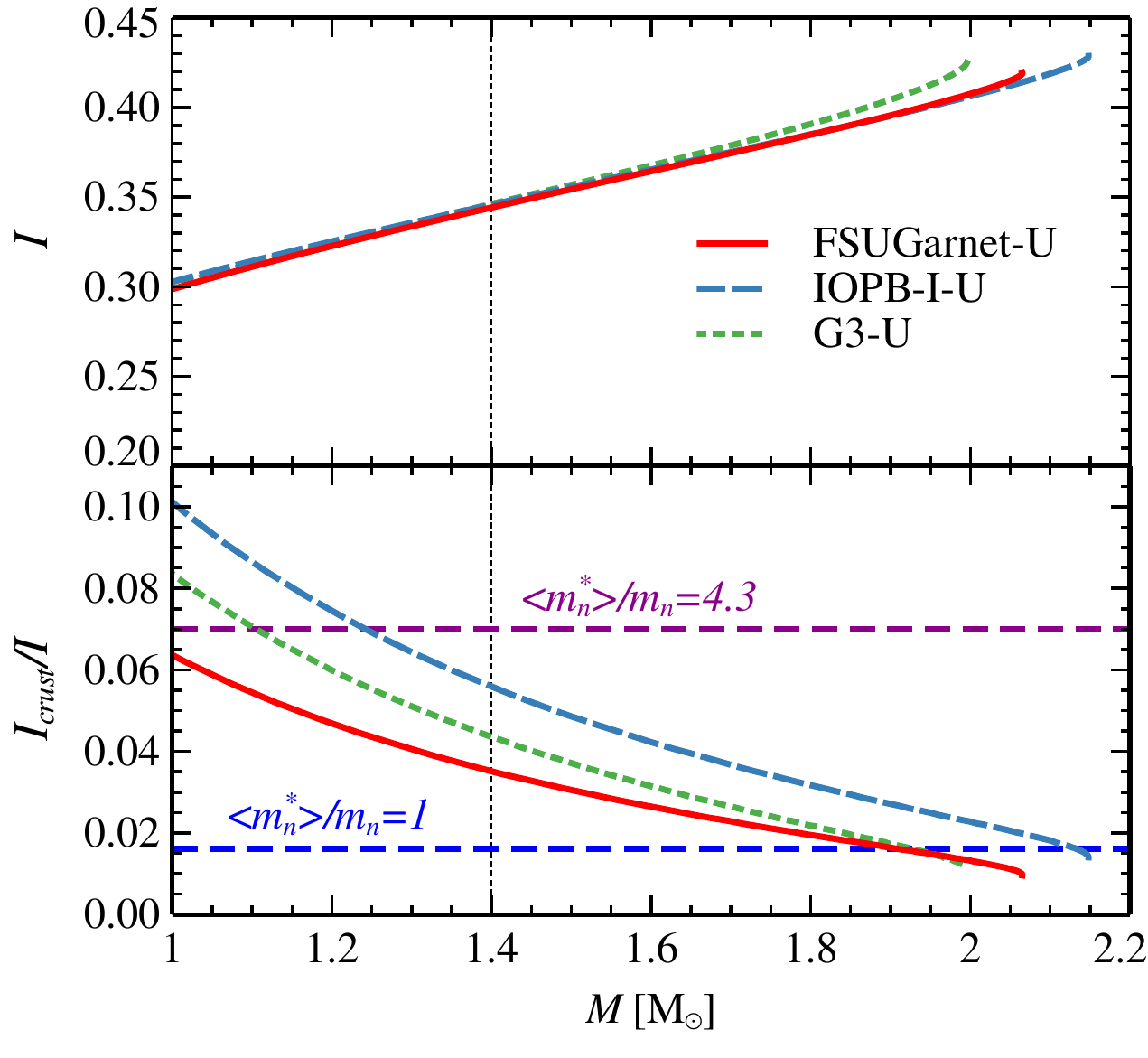}
    \caption{{\it Upper:} The normalized moment of inertia as a function of mass for three unified EoSs. {\it Lower:} The fractional moment of inertia as a function of mass. The dashed dark magenta and dark blue lines represent the Vela pulsar data (see text for details).}
    \label{fig:mom}
\end{figure}

The moment of inertia of the neutron star is calculated for a uniformly rotating case (slow rotation) as described in Sub-Sec. \ref{form:NS_observable}. The total normalized MI of the neutron star is shown in the upper panel of Fig. \ref{fig:mom} for three unified EoSs. The $I$ increases with the masses of the neutron star as it depends on the mass of the star. The $I$ for considered sets is almost  same up to $1.6 \ M_\odot$ and then slightly diverges. This is because the core part of EoS is  model-dependent. Some theoretical predictions believe that the relation between $I$ and $M$ is universal \cite{Lattimer_2005, LATTIMER_2016, Landry_2018}. It means that one can predict the nature of $I$ from the observed mass of the star. 

The crustal MI of the neutron star is calculated using Eq. (\ref{eq:moic}) from the crust-core transition radius $R_c$ to the surface of the star $R$. The fractional moment of inertia ($I_{crust}/I$) is depicted in the lower panel of Fig. \ref{fig:mom}. It is seen that for a massive  neutron star, the lesser moment of inertia is stored in the crust. In this case, the maximum mass, FMI for the canonical star, FMI$_{1.4}$ predicted by IOPB-I-U EoS is $2.149\ M_\odot$ and $\approx 0.057$ respectively. For FSUGarnet-U and G3-U cases, the masses and FMI$_{1.4}$ are ($2.065 \ M_\odot$, 0.044) and ($1.996 \ M_\odot$, 0.036) respectively as given in Table \ref{tab:NS_observables}. The blue and violet dashed lines represent the minimum value needed to justify the Vela glitch with \cite{Andersson_2012} and without \cite{Link_1999} crustal entrainment. The details on the crustal entrainment are discussed in the following subsection. It is evident that the crustal moment of inertia is sensitive to the crust's mass and radius, which subsequently depends on the crust-core transition density and the pressure. Therefore accurate estimation of these properties is an essential and unified treatment of EoS become pivotal.

\begin{table*}
\caption{The neutron star properties such as maximum mass ($M_{\rm max}$), maximum radius ($R_{max}$), canonical radius ($R_{1.4}$), normalized maximum MI ($I_{max}$), normalized canonical MI ($I_{1.4}$), maximum FMI (FMI$_{max}$), canonical FMI (FMI$_{1.4}$), mass of the crust ($M_{\rm crust}$), and length of the crust ($l_{crust}$) for FSUGarnet, IOPB-I, and G3 EoSs.}
\label{tab:NS_observables}
\begin{tabular}{llllllllll}
\hline \hline
EoSs &
  \begin{tabular}[c]{@{}l@{}}$M_{\rm max}$\\ ($M_\odot$)\end{tabular} &
  \begin{tabular}[c]{@{}l@{}}$R_{\rm max}$\\ (km)\end{tabular} &
  \begin{tabular}[c]{@{}l@{}}$R_{1.4}$\\   (km)\end{tabular} &
  $I_{\rm max}$ &
  $I_{1.4}$ &
  FMI$_{\rm max}$ &
  FMI$_{1.4}$ &
  \begin{tabular}[c]{@{}l@{}} $M_{\rm crust}$\\ ($M_\odot$)\end{tabular} &
  \begin{tabular}[c]{@{}l@{}}$l_{\rm crust}$\\(km)\end{tabular}  \\ \hline
 IOPB-I-U & 2.148 & 11.947 & 13.301 & 0.429 & 0.346 & 0.014 & 0.057 & 0.013  & 0.490 \\ \hline
 FSUGarnet-U& 2.065 & 11.775 & 13.170  & 0.419  & 0.344 & 0.010 & 0.044 &0.009 &0.528  \\ \hline
 G3-U& 1.996  &10.942 & 12.598 & 0.426  & 0.346 & 0.011 & 0.036  & 0.010 & 0.451  \\ \hline \hline
\end{tabular}
\end{table*}
\subsection{Pulsar glitch }
\label{res:pulsar_glitch}

\begin{figure*}[ht]
    \centering
    \includegraphics[width = 0.7\textwidth]{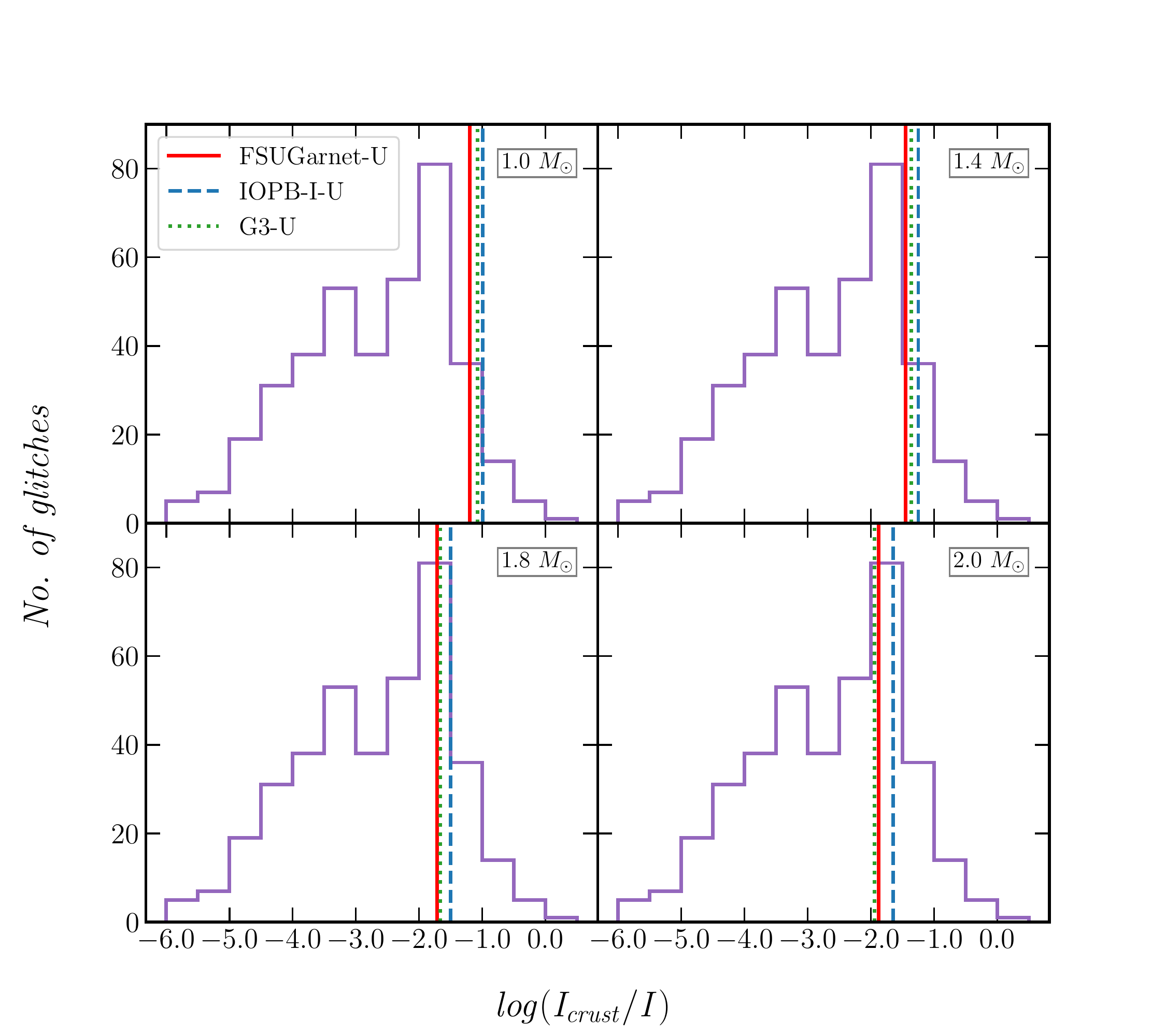}
    \caption{Distribution of $I_{crust}/I$ calculated using 581 glitches \cite{Espionza_2011}. The vertical lines are the FMI for FSUGarnet-U, IOPB-I-U, and G3-U EoSs.}
    \label{fig:glitch}
\end{figure*}

Pulsars are rotating neutron stars observed to have pulses of radiation at very regular intervals that typically range from milliseconds to seconds. Pulsars have very strong magnetic fields which funnel jets of particles out along the two magnetic poles. These accelerated particles produce very powerful beams of light. The pulsed emission, which is in the radio frequency band, is the direct way of measuring the rotation of the crust using the pulsar timing technique \cite{Basu_2018}. By measuring the time of arrival of the pulse, one can estimate the crust's rotational speed and glitch activity.

The glitches are produced due to the sudden spin-ups in the radio pulsars. This is because the angular momentum transfers from the superfluid component of the stellar interior to the solid crust. Therefore, there is a change of MI from the superfluid to the rest of the star. The fractional crustal moment of inertia (FMI) is the ratio of the total MI to the crustal MI ($I_{crust}/I$), and it is related to the characteristic pulsar glitches properties \cite{Basu_2018, Eya_2017},
\begin{equation}
    \frac{I_{crust}}{I}=2\tau_c\frac{1}{t_i}\Big(\frac{\Delta \nu}{\nu}\Big)_i,
\label{eq:glitch}
\end{equation}
where $\tau_c$ is the characteristic age of the pulsar, $t_i$ is the time elapsed before the $i$th glitch since the preceding glitch and $\Big(\frac{\Delta \nu}{\nu}\Big)_i$ is fractional frequency jump. From the above relation, one can compare the theoretical FMI with the observational results. 

Inside the neutron star, the neutron superﬂuid is strongly coupled to the solid crust due to nondissipative entrainment eﬀects \cite{Chamel_2013, G_gercino_lu_2014}. These effects limit the amount of angular momentum that can be transferred during a glitch event. The importance of the entrainment coupling is related to the neutron eﬀective mass $m^*_n$ in the inner crust, which is proportional to the ratio of unbound neutrons to those that are not entrained \cite{carreau2020modeling}. With this entrainment eﬀects, the Eq. (\ref{eq:glitch}) can be written as 
\begin{eqnarray}
 \frac{I_{crust}}{I}=2\tau_c\frac{\left<m_n^*\right>}{m_n}\frac{1}{t_i}\Big(\frac{\Delta \nu}{\nu}\Big)_i,
\label{eq:glitch_ent}
\end{eqnarray}
where $\left<m_n^*\right>$ is the average effective mass of neutrons in the inner crust. The ratio of the $\left<m_n^*\right>/m_n$ has value 4.3 \cite{Andersson_2012} and the ratio becomes one ($\left<m_n^*\right>=m_n$) where no crustal entrainment are considered \cite{Link_1999}. 


In Fig. \ref{fig:glitch}, we plot the FMI estimated from the observed 581  \footnote{{\bf \url{http://www.jb.man.ac.uk/pulsar/glitches.html}}} glitches catalogue \cite{Espionza_2011}. With addition to this, we calculate the theoretical FMI using Eqs. (\ref{eq:moic} and \ref{eq:moi}) for three unified EoSs with different masses of the star. The FMIs for theoretical calculations are well consistent with peak in case for $1.8\ M_\odot$ and $2.0\ M_\odot$ masses. 
\section{Conclusion}
\label{conclusion}
In summary, we provide the unified treatment of EoS of the neutron star, namely FSUGarnet-U, IOPB-I-U, and G3-U. We consider different physics for various layers beginning from the outer crust to the inner core within the E-RMF framework. The outer crust is treated within the well-known variational BPS formalism, while the structure of the inner crust is calculated using the compressible liquid drop model.  We use the most recent atomic mass evaluation AME2020 and the highly precise microscopic HFB mass models along with the experimental mass of available neutron-rich nuclei to find the equilibrium composition of the outer crust.  We compare the EoS and composition of outer crust calculated from different mass models and find the persistent existence of $Z=28$ and $N= 50$ and $82$ nuclei. The majority of mass models predict the presence of even mass nuclei in the outer crust except for the HFB-14, which indicate a thin layer of $^{121}$Y at high pressure suggesting a possible ferromagnetic behavior of neutron star. 

The inner crust is treated with the CLDM formalism using the E-RMF framework to calculate the bulk and finite-size energies of the cluster. The composition of the inner crust using the CLDM is in harmony with the available microscopic predictions. The  mass, asymmetry, and gas density increase monotonically with  baryon density or star's depth while the cluster becomes dilute. It is seen that the equilibrium configuration of the inner crust is strictly model-dependent and depends mainly on symmetry energy and slope parameter  in the subsaturation density regime, and the surface energy parametrization. A higher value of symmetry energy or lower slope parameter results in the larger  mass and charge of the cluster. We also calculate the crust-core transition density ($\rho_t$) and pressure ($P_t$) from the crust side and find that these values are sensitive to the isovector surface parameter $p$ and slope parameter $L$. The values of $\rho_t$ for the FSUGarnet-U, IOPB-I-U, and G3-U are found to be 0.08755, 0.07114, and 0.08125 fm$^{-3}$ whereas the $P_t$ is calculated as 0.46793, 0.31415 and 0.45284 MeV fm$^{-3}$ respectively.  In addition, we also show the behavior of adiabatic index, shear, and compression modulus in the inner crust region. 
The neutron star properties such as mass, radius, and the moment of inertia are calculated with three unified EoSs viz. FSUGarnet-U, IOPB-I-U, and G3-U. The masses predicted by the three EoSs are well consistent with the different massive pulsars data. The predicted canonical radii are well within  the old and NICER constraints limits. The crustal mass and thickness are also calculated with three unified EoSs. We observe that the crust is thicker for low mass neutron star, and it drops continuously with increasing neutron star mass.

The moment of inertia is calculated for a slowly rotating neutron star. The MI increases with increasing the star's mass, and it is almost unchanged around $1.6\ M_\odot$, then it diverges. From the theoretical predictions, it is believed that there exist some Universal relations between MI and mass of the neutron star. In  future, we expect that more pulsars detection (glitch events) and binary neutron star merger events may put  tight constraints on the MI.

The pulsars are rotating neutron stars, which emit the electromagnetic frequency with millisecond time intervals. This is because the glitches are produced due to the sudden spin-ups in the radio frequency due to the angular momentum transfer from the superfluid part to the outer crust. To illustrate the glitch event, we calculate the FMI for three EoSs. We observed that the more massive a neutron star is, the less MI stores in its crust. We constraint the FMI by putting Vela pulsars data with and without entrainment of the crust. We compare the FMI from the theoretical with observed data approximately for 581 glitches. The theoretical prediction is well consistent with the highest peak for canonical to maximum mass star. This implies that the maximum number of glitches observed so far are well compatible with our theoretical results. 

In this work, we restrict ourselves to the spherically symmetric Wigner-Seitz cell as nonspherical structures do not affect the EoS significantly. However, the existence of nonspherical structures  close to the crust-core interface have various observational consequences. Therefore to access the impact of pasta structures, we shall perform a comprehensive analysis of neutron star crust including nonspherical shapes in the future work.

In conclusion, we summarized that the three unified EoSs, FSUGarnet-U, IOPB-I-U, and G3-U, well reproduced the observational data obtained with different pulsars, NICER, and glitch. Hence, these  unified EoSs may be used for future exploration of more neutron star properties such as transport, cooling, inspiral etc. 
\section*{Acknowledgment}
One of the authors, V.P., would like to thank the support provided by the Institute of Physics Bhubaneswar during the completion of this work. V.P. also thank Thomas Carrareu for the discussion on the code of CLDM formulation.
\bibliography{crust}
\end{document}